\documentclass[fleqn,usenatbib]{mnras}
\usepackage{aas_macros}
\usepackage{abbreviations}
\usepackage{amssymb}
\usepackage{bigdelim}
\usepackage{bm}
\usepackage{cprotect}
\usepackage{etoolbox}
\usepackage{gensymb}
\usepackage{graphicx}
\usepackage{hyperref}
\usepackage{mathtools}
\usepackage{multirow}
\usepackage{multicol}
\usepackage{subcaption}
\usepackage{threeparttable}
\usepackage{upgreek}
\usepackage{xcolor}
\usepackage[figuresright]{rotating}
\usepackage{newtxtext,newtxmath}
\usepackage[T1]{fontenc}
\usepackage{orcidlink}

\definecolor{applegreen}{rgb}{0.55, 0.71, 0.0}
\def\xmm{\textit{XMM-Newton}}
\def\chandra{\textit{Chandra}}

\hypersetup{
  colorlinks   = true, 
  urlcolor     = blue, 
  linkcolor    = red, 
  citecolor   = blue 
}

\title[MeerKAT-meets-LOFAR: Abell~1413]{A MeerKAT-meets-LOFAR study of Abell~1413: a moderately disturbed non-cool-core cluster hosting a $\bm{\sim}$500~kpc `mini'-halo}

\author[C.~J.~Riseley et al.]{C.~J.~Riseley\,\orcidlink{0000-0002-3369-1085}\,$^{1,2,3}$\thanks{Corresponding author email: \url{christopher.riseley@unibo.it}}, \hspace{0.001cm} N.~Biava\,\orcidlink{0000-0001-7947-6447}\,$^{1,2}$, G.~Lusetti\,\orcidlink{0009-0009-1363-4671}\,$^{4}$,
A.~Bonafede\,\orcidlink{0000-0002-5068-4581}\,$^{1,2}$, 
E.~Bonnassieux\,\orcidlink{0000-0003-2312-3508}\,$^{5,6,2}$, 
A.~Botteon\,\orcidlink{0000-0002-9325-1567}\,$^2$, 
\newauthor
F.~Loi\,\orcidlink{0000-0002-8627-6627}\,$^7$, G.~Brunetti\,\orcidlink{0000-0003-4195-8613}\,$^2$, R.~Cassano\,\orcidlink{0000-0003-4046-0637}\,$^2$, E.~Osinga\,\orcidlink{0000-0002-5815-8965}\,$^8$, K.~Rajpurohit\,\orcidlink{0000-0001-7509-2972}\,$^{9,1,2}$, H.~J.~A.~R\"{o}ttgering\,\orcidlink{0000-0001-8887-2257}\,$^8$
\newauthor
T.~Shimwell$^8$, R.~Timmerman\,\orcidlink{0000-0001-9404-2612}\,$^8$, R.~J.~van Weeren\,\orcidlink{0000-0002-0587-1660}\,$^8$ \\
$^1$ Dipartimento di Fisica e Astronomia, Universit\`a degli Studi di Bologna, via P. Gobetti 93/2, 40129 Bologna, Italy \\ 
$^2$ INAF -- Istituto di Radioastronomia, via P. Gobetti 101, 40129 Bologna, Italy \\ 
$^3$ CSIRO Space \& Astronomy, PO Box 1130, Bentley, WA 6102, Australia \\ 
$^4$ Hamburger Sternwarte, Universit\"{a}t Hamburg, Gojenbergsweg 112, 21029 Hamburg, Germany \\
$^5$ Julius-Maximilians-Universit\"{a}t W\"{u}rzburg, Fakult\"{a}t f\"{u}r Physik und Astronomie, Institut f\"{u}r Theoretische Physik und Astrophysik, Lehrstuhl f\"{u}r Astronomie, \\
\hspace{0.1cm} Emil-Fischer-Str. 31, D-97074 W\"{u}rzburg, Germany \\
$^6$ Observatoire de Paris, 5 place Jules Janssen, 92195 Meudon, France \\
$^7$ INAF-Osservatorio Astronomico di Cagliari, Via della Scienza 5, 09047 Selargius, CA, Italy \\ 
$^8$ Leiden Observatory, Leiden University, P.O. Box 9513, 2300 RA Leiden, The Netherlands \\ 
$^9$ Harvard-Smithsonian Center for Astrophysics, 60 Garden Street, Cambridge, MA 02138, USA
}

\date{Accepted 2023 July 20. Received 2023 July 13; in original form 2023 May 31.}

\pubyear{2023}

\begin{document}
\label{firstpage}
\pagerange{\pageref{firstpage}--\pageref{lastpage}}
\maketitle

\begin{abstract}
Many relaxed cool-core clusters host diffuse radio emission on scales of hundreds of kiloparsecs: mini-haloes. However, the mechanism responsible for generating them, as well as their connection with central active galactic nuclei, is elusive and many questions related to their physical properties and origins remain unanswered. This paper presents new radio observations of the galaxy cluster Abell 1413 performed with MeerKAT (L-band; 872 to 1712 MHz) and LOFAR HBA (120 to 168 MHz) as part of a statistical and homogeneous census of mini-haloes. Abell 1413 is unique among mini-halo clusters as it is a moderately-disturbed non-cool-core cluster. Our study reveals an asymmetric mini-halo up to 584~kpc in size at 1283~MHz, twice as large as first reported at similar frequencies. The spectral index is flatter than previously reported, with an integrated value of $\alpha = -1.01 \pm 0.06$, shows significant spatial variation, and a tentative radial steepening. We studied the point-to-point X-ray/radio surface brightness correlation to investigate the thermal/non-thermal connection: our results show a strong connection between these components, with a super-linear slope of $b = 1.63^{+0.10}_{-0.10}$ at 1283 MHz and $b = 1.20^{+0.13}_{-0.11}$ at 145 MHz. We also explore the X-ray surface brightness/radio spectral index correlation, finding a slope of $b = 0.59^{+0.11}_{-0.11}$. Both investigations support the evidence of spectral steepening. Finally, in the context of understanding the particle acceleration mechanism, we present a simple theoretical model which demonstrates that hybrid scenarios — secondary electrons (re-)accelerated by turbulence — reproduce a super-linear correlation slope.
\end{abstract}


\section{Introduction}
Diffuse radio sources in clusters of galaxies trace the presence of non-thermal components -- magnetic fields and relativistic electrons (or cosmic ray electrons, hereafter CRe) -- on some of the largest scales in the Universe. Broadly speaking, these sources can be categorised into three principal classes dating back around two decades \citep[e.g.][]{Kempner2004}: radio relics (also known as radio shocks or `radio gischt'), radio haloes, and radio mini-haloes (hereafter referred to simply as mini-haloes). As of 2019, a few hundred galaxy clusters had been detected which host some combination of radio relic(s) and/or a radio halo \citep[for the most recent observational review of these sources, see][]{vanWeeren2019}. 

With the advent of next-generation instrumentation such as the LOw-Frequency ARray \citep[LOFAR;][]{vanHaarlem2013}, the Murchison Widefield Array \citep[MWA;][]{Tingay2013}, and the MeerKAT telescope (\citealt{Jonas2016}, though see also \citealt{Camilo2018} and \citealt{Mauch2020} for discussion of MeerKAT's technical capabilities), the number of diffuse radio sources associated with clusters of galaxies is increasing rapidly \citep[e.g.][]{Wilber2018,Wilber2020,Locatelli2020,Hoang2021,Hoeft2021,Botteon2018,Botteon2019,Botteon2022,Duchesne2020,Duchesne2021a,Duchesne2021b,Duchesne2021c,Duchesne2022,Biava2021_RXCJ1720,Knowles2022,Venturi2022,Hoang2022,Riseley2022_A3266}.

Mini-haloes are perhaps the most elusive of the three canonical classes of diffuse radio source, with only some 35 currently catalogued, compared to some hundreds of relics and haloes. They are moderately-extended radio sources, typically around $0.1-0.3$~Mpc in size and are predominantly hosted by relaxed clusters. While relatively few mini-haloes have been catalogued to-date, statistical studies suggest that their occurrence rates are high among cool-core clusters, up to $\sim80\%$ for massive clusters with $M_{\rm 500} \gtrsim 6 \times 10^{14} \, M_{\odot}$, although there is tentative evidence for a decreased rate among lower mass cool-core clusters \citep[see discussion by][]{Giacintucci2017}.

The known population of mini-halo clusters is distributed across a broad range of redshifts from $z \sim 0.01$ to $z \sim 0.81$ \citep[see][and references therein]{vanWeeren2019}. Due to the inefficient nature of the acceleration mechanism(s) responsible, relics, haloes and mini-haloes all possess a steep radio spectrum, nominally $\alpha \lesssim -1$ (where $\alpha$ is the spectral index, relating flux density $S$ to observing frequency $\nu$ as $S \propto \nu^{\alpha}$).

In the literature, there has been much discussion around the particle acceleration mechanism(s) that power mini-haloes. Broadly speaking, these fall into two classes: primary models \citep[also known as the `turbulent (re-)acceleration' scenario;][]{Gitti2002,ZuHone2013} and secondary models \citep[also known as the `hadronic scenario';][]{Pfrommer2004}. For a review of all scenarios, see also \cite{BrunettiJones2014}.

Under the primary/turbulence scenario, electrons are accelerated to the relativistic regime (GeV energies) by cluster-scale magnetohydrodynamic (MHD) turbulence that is generated by core sloshing. In relaxed clusters, such core sloshing could be induced by dynamical interactions and/or minor/off-axis mergers with other clusters/groups. One of the most striking pieces of observational evidence for core sloshing comes from the detection of cold fronts and/or large-scale spiral motions in X-ray observations \citep{Mazzotta2001,Mazzotta2001b,Markevitch2007,Mazzotta2008,Owers2009,Ghizzardi2010,Ghizzardi2014,Johnson2012,PaternoMahler2013,Rossetti2013,Savini2018,Botteon2018_ChandraProfiles,Riseley2022_MS1455}. Such sloshing motions have also been investigated from the theoretical perspective \citep[e.g.][]{Ascasibar2006,ZuHone2013}.

In general, simulations have shown that these events can replicate several of the observed properties of clusters hosting mini-haloes, including large-scale bulk motions required to generate sloshing spirals \citep{ZuHone2013,Machado2015} and fluctuations in radio spectral index that might be expected to arise from inhomogeneities in turbulence \citep{Giacintucci2014b}. In recent years, much work has been devoted to exploring potential sources for the seed electrons. The co-location of many mini-haloes with a central radio brightest cluster galaxy (BCG) suggests a natural source for this population, and emerging correlations between the radio power of BCGs and mini-haloes, as well as mini-halo power and X-ray cavity power, provide observational support \citep{Bravi2016,RichardLaferriere2020}. While the BCG is commonly held to be the dominant factor in energy input into the ICM, recent work by \cite{Seth2022} has suggested that non-central radio galaxies may also provide a significant source of energy input into the ICM.

According to the secondary/hadronic scenario, CRe are continuously injected into the ICM through collisions between cosmic ray protons (CRp) and thermal protons. One of the natural sources of CRp in clusters are active galactic nuclei (AGN); due to their longer radiative lifetime, CRp are expected to persist throughout much of the cluster volume \citep{BrunettiJones2014}. 

Both scenarios share a number of commonalities, including (i) the important role of AGN, in particular the central BCG, and (ii) the connection between non-thermal and thermal components in the ICM. As such, investigating correlations between a number of observables --- such as radio/X-ray surface brightness --- is a key tool with which we can probe the underlying acceleration mechanism \citep[e.g.][]{Govoni2001}.

With the established suite of Square Kilometre Array (SKA) Pathfinder and Precursor instruments, a number of recent works have begun to explore these correlations in increasing detail \citep{Ignesti2020,Ignesti2021,Biava2021_RXCJ1720,Timmerman2021,Riseley2022_MS1455,Riseley2022_A3266}. In particular \cite{Ignesti2020} studied point-to-point correlations for a first statistical sample of seven mini-haloes. The point-to-point correlation between radio surface brightness and X-ray surface brightness takes the form ${\rm log}(I_{\rm{R}}) \propto b \, {\rm log}(I_{\rm{X}})$. 

Giant radio halos are generally believed to be generated by cluster-scale turbulence, and show a sub-linear correlation slope (i.e. $b < 1$), which arises because both particles (acceleration and transport) and fields (amplification by turbulent-dynamo) follow the spatial distribution of turbulence, which is very broad. For point-to-point studies of radio haloes, see for example: \cite{Govoni2001,Brown&Rudnick2011,Botteon2020b_Abell2255,Xie2020,Bruno2021,Duchesne2021b,Rajpurohit2018,Rajpurohit2021_MACSJ0717_Halo,Rajpurohit2021_A2744,Hoang2019,Hoang2021,Bonafede2022,Vacca2011,Vacca2022,Riseley2022_A3266,Botteon2023_HaloEdges}. 

However, \cite{Ignesti2020} found a typically super-linear correlation slope (i.e. $b > 1$) for their mini-halo sample. While a super-linear correlation slope arises naturally under the secondary/hadronic scenario, the primary/turbulence scenario can also replicate a super-linear correlation slope if either the CRe profile and/or the turbulence strength is more peaked toward the cluster centre. This would, however, suggest that mini-haloes trace a turbulent component that is different from the one powering giant halos. As such, neither scenario is completely ruled out through point-to-point correlations alone.

Complicating our phenomenology, a growing number of mini-halo clusters show diffuse radio emission far outside the cool core, and far beyond any cold fronts/sloshing spiral structures \citep[][Biava et al., in prep.]{Venturi2017,Savini2018,Savini2019,Biava2021_RXCJ1720,Riseley2022_MS1455}. This challenges our theoretical understanding of particle acceleration mechanisms, which are typically confined to the sloshing regions \citep{ZuHone2013,ZuHone2015}. 

One possibility is that hadronic interactions play an important role within the cool core, while turbulence becomes progressively dominant on larger scales \citep[e.g.][]{Cassano2012,Zandanel2014}. The steep spectrum observed for diffuse emission outside the cool core in a handful of cases favours a turbulent acceleration interpretation \citep{Savini2019,Biava2021_RXCJ1720}.

\begin{figure}
\begin{center}
\includegraphics[width=\linewidth]{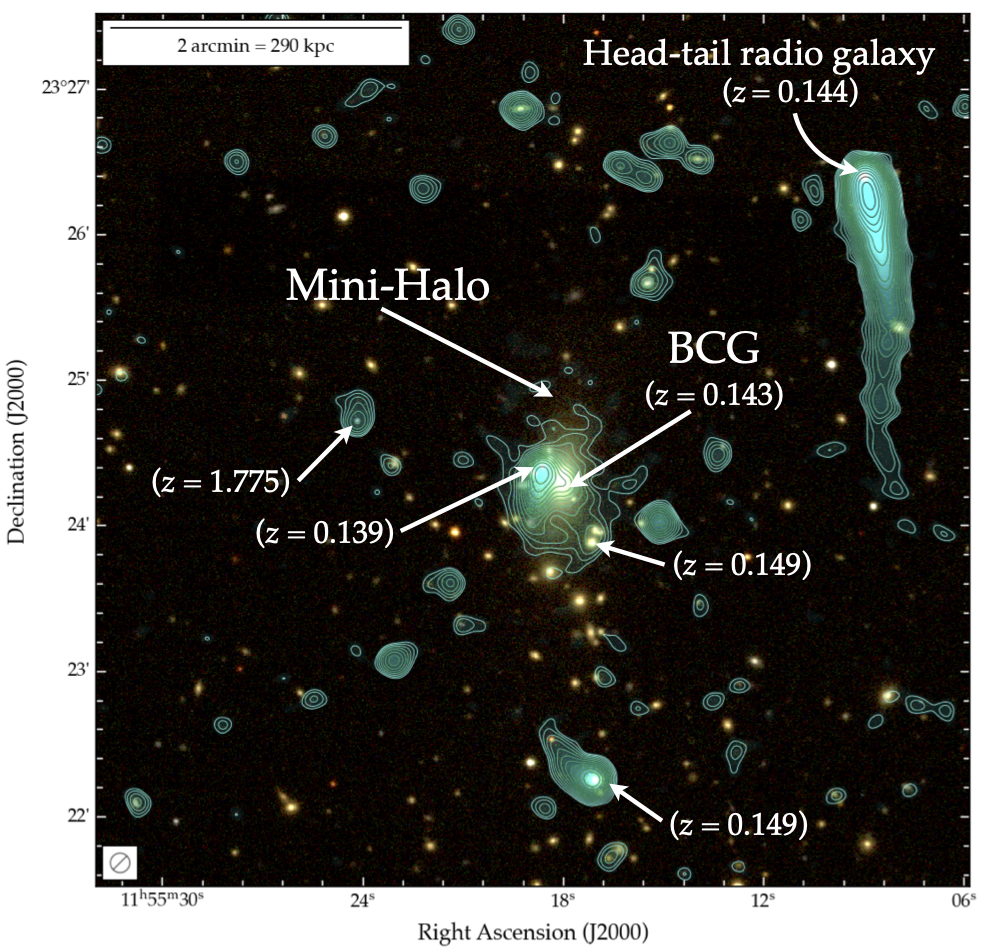}
\cprotect\caption{Colour-composite image of Abell~1413. The teal colourmap and contours represent the 1283~MHz radio surface brightness measured by MeerKAT at $8$~arcsec resolution (teal color and contours). Contours start at $4\sigma$ and scale by a factor $\sqrt{2}$, where $\sigma = 6.6~\upmu$Jy beam$^{-1}$; see Table~\ref{tab:img_summary}. The radio surface brightness is overlaid on an optical RGB image constructed using $i$-, $r$- and $g$-band images from SDSS DR16 \citep{Ahumada2020}. The redshifts of selected galaxies are also shown for reference.}
\label{fig:Abell1413_rgb}
\end{center}
\end{figure}

\subsection{The MeerKAT-meets-LOFAR mini-halo census}
We are carrying out a census of all known radio mini-haloes in the Declination range $-1\degree$ to $+30\degree$ to determine the nature of particle acceleration mechanisms at play in mini-haloes using the first uniformly-constructed mini-halo sample. The non-thermal window into our sample is provided by deep MeerKAT observations covering the frequency range 872$-$1712~MHz, and LOFAR HBA observations covering the frequency range 120$-$168~MHz. MeerKAT observations are being carried out under MeerKAT Project ID (PID) SCI-20210212-CR-01 (P.I. Riseley). LOFAR observations are sourced primarily from the LOFAR Two-metre Sky Survey \citep[LoTSS;][]{Shimwell2017_LoTSS_PaperI,Shimwell2019_LOTSS_PaperII,Shimwell2022_LOTSS_PaperIII} and expanded with targeted observations where LoTSS coverage is not yet completed. This multi-frequency radio dataset is augmented by archival X-ray data from \textit{Chandra} and \xmm{} to provide insights into the thermal properties of the mini-halo sample. 

The first paper from this census detailed observations of the galaxy cluster MS~1455.0$+$2232 \citep{Riseley2022_MS1455}. To summarise the findings, we reported (i) the detection of significant diffuse radio emission on linear scales up to 586~kpc, far larger than previously reported, (ii) the detection of a sloshing spiral 254~kpc in extent, seen in the \chandra{} X-ray surface brightness gradient map, and (iii) a consistent super-linear slope in the point-to-point radio/X-ray surface brightness correlation. While both the primary/turbulence and secondary/hadronic scenarios were able to explain some of the observational properties of MS~1455.0$+$2232, neither could provide a fully satisfactory explanation of all the observed properties.

\subsection{The galaxy cluster Abell~1413}
In this paper, we report on another galaxy cluster selected for study as part of our census: Abell~1413. Figure~\ref{fig:Abell1413_rgb} shows a colour-composite image of this cluster, with the 1283~MHz radio surface brightness measured by MeerKAT overlaid on an optical RGB image constructed using \textit{i-}, \textit{r-}, and \textit{g-}band images from Data Release 16 of the Sloan Digital Sky Survey \citep[SDSS DR16;][]{Ahumada2020}.

Abell~1413 is an intriguing galaxy cluster at redshift $z = 0.143$, with a mass of $M_{500} = (5.98^{+0.48}_{-0.40}) \times 10^{14}~{\rm M}_{\odot}$ \citep{Planck2014}. This cluster has been the target of varied cosmological and galaxy evolution studies using multi-wavelength data \citep[e.g.][]{Grainge2002,Morrison2003}. At (sub-)mm wavelengths, observations of the Sunyaev-Zeldovich (SZ) effect toward Abell~1413 suggest that the cluster exhibits an overall relaxed morphology \citep{Grainge1996,Bonamente2006,LaRoque2006,AMI2012}, although there is a noticeable offset between the peak of the SZ signal and the peak of the X-ray emission corresponding to the thermal ICM \citep[see discussion in][]{AMI2012}.

\cite{Castagne2012} present a detailed optical analysis of the cluster-member galaxy population using data from SDSS DR7 \citep{SDSS-DR7} and new observations from the Canada-France-Hawaii Telescope. While the overall galaxy velocity distribution suggests a relaxed morphology, their analysis revealed several substructures within Abell~1413 that are broadly aligned along a north-south axis (see \citealt{Castagne2012} for details).

At X-ray wavelengths, Abell~1413 has been studied extensively with various X-ray missions, including \xmm{} and \textit{Chandra} \citep[][Lusetti et al., in prep.]{Pratt2002,Pointecouteau2005,Vikhlinin2005,Baldi2007,Snowden2008,Hoshino2010,Ettori2010,Bartalucci2017,Botteon2018_ChandraProfiles,Lusetti_Tesi2021}. Overall, Abell~1413 presents a slightly disturbed X-ray morphology, elongated in the north-south direction, although there is no strong evidence of a recent merger. The cluster has a higher central temperature $(T_{\rm X} = 8.3 \pm 0.2~{\rm keV})$ and higher central entropy (${\rm K_0} = 64 \pm 8$~keV~cm$^{2}$) than expected for typical relaxed clusters \citep{Giacintucci2017,Botteon2018_ChandraProfiles}.

However, Abell~1413 shows a low centroid shift $w = 0.04^{+0.01}_{-0.02}$ and a moderate concentration parameter, $c = 0.44 \pm 0.04$ \citep[both][]{Campitiello2022_CHEXMATE} that are more typical of relaxed. We note that the definition of $c$ used by \cite{Campitiello2022_CHEXMATE} differs from that adopted by e.g. \cite{Rossetti2017}, who find $c = 0.102 \pm 0.001$. However, given the respective definitions of these metrics (see the discussions by those authors) the interpretation is the same: Abell~1413 overall shows a mixed X-ray morphology, neither fully relaxed nor disturbed.

Despite the lack of a an identified cool core, Abell~1413 hosts a known mini-halo first identified as a candidate by \cite{Govoni2009} and later confirmed by \cite{Savini2019} using data from LOFAR High Band Antennas (HBA) at 145~MHz. The lack of a cool core is highly unusual for mini-haloes: in the sample of \cite{Giacintucci2017}, Abell~1413 is the only non-cool-core cluster to host a mini-halo. However, mini-haloes hosted by non-cool-core clusters may be more common than this. Recent work from the MeerKAT Galaxy Cluster Legacy Survey \citep[MGCLS;][]{Knowles2022} reported the detection of seven clusters hosting newly-detected candidate mini-haloes and three confirmed new mini-haloes, including  Abell~4038 and MCXC~J0342.8$-$5338 (Abell~3158). X-ray observations of these clusters reveal that neither shows typical characteristics of relaxed cool-core clusters. See respectively \cite{Rossetti_Molendi_2010} and \cite{Whelan_2022} and references therein for discussion.

The `mini'-halo has also recently been studied using more recent LOFAR observations by \cite{Lusetti_Tesi2021} and Lusetti et al. (in prep.), who found that the diffuse emission covers a larger physical scale at 145~MHz than previously thought, potentially up to $\sim800$~kpc in extent, and exhibits evidence of multiple components. These authors adopt the interpretation that Abell~1413 hosts a ``mini-halo-plus-giant-halo'' type structure, although in this paper we will continue to refer to the entire diffuse source as a `mini'-halo. 

The radio counterpart to the BCG of Abell~1413 was only recently reported for the first time by \cite{Savini2019}. Previous higher-frequency observations at 1.4~GHz achieved insufficient sensitivity to detect this faint source \citep{Govoni2001,Giacintucci2017}. \cite{RichardLaferriere2020} include Abell~1413 in their sample of mini-haloes from which they re-derive scaling relations between mini-halo radio power and BCG radio power; however, with the available high-frequency radio data, they were only able to place limits on the radio power of the BCG at GHz frequencies. Finally, Abell~1413 was also recently studied as part of a mini-sample by \cite{Trehaeven2023} using relatively shallow MeerKAT observations. They report detections of both the radio counterpart to the BCG as well as the embedded head-tail radio galaxy at 1.28~GHz, as well as a measuring a linear size up to 211~kpc for the mini-halo. They also study the in-band spectral properties for these three sources, finding an ultra-steep in-band spectral index of $\alpha = -1.52 \pm 0.46$, although the uncertainty is large.

In this paper, we present new deep observations across the radio spectrum from 145~MHz to 1.7~GHz, taken with MeerKAT, the upgraded Giant Metrewave Radio Telescope \citep[uGMRT;][]{Gupta2017} and LOFAR HBA. The remainder of this paper is divided as follows: we discuss the observations and data reduction in \S\ref{sec:observations}, we present our results in \S\ref{sec:results} and analyse them in \S\ref{sec:analysis}. We draw our conclusions in \S\ref{sec:conclusions}. Throughout, we assume a $\Lambda$CDM cosmology of H$_0 = 73 ~ \rm{km} ~ \rm{s}^{-1} ~ \rm{Mpc}^{-1}$, $\Omega_{\rm{m}} = 0.27$, $\Omega_{\Lambda} = 0.73$. At the representative redshift of Abell~1413 \citep[$z = 0.143$;][]{Sanders2011} the angular scale to linear size conversion is 1~arcsec to 2.417~kpc, with our cosmology. We quote all uncertainties at the $1\upsigma$ level.

\section{Observations \& Data Reduction}\label{sec:observations}

\subsection{Radio: MeerKAT}
Abell~1413 was observed with the MeerKAT telescope on two separate occasions: 2019 June 06 (CBID\footnote{Capture Block ID}~1565438457) and 2021 March 24 (CBID~1631864177). The 2019 observation was carried out under the `Mining Minihalos with MeerKAT' project \citep[see][]{Trehaeven2023}, whereas the 2021 observation was carried out under our project, SCI-20210212-CR-01. Both observing runs were performed using the L-band receiver system, with 4096 channels covering the frequency range 872$-$1712~MHz. 

The bandpass calibrator PKS~B0407$-$658 was observed for ten minutes at the beginning of each observing run; to track the time-varying instrumental gains, the compact radio source J1120$+$1420 was observed for two minutes at a quarter-hour cadence (CBID~1565438457) or half-hour cadence (CBID~1631864177). While the data from CBID~1565438457 have very recently been published by \cite{Trehaeven2023}, our processing occurred independently, including our more recent observations which provided an overall increase in on-source time from 113~minutes (CBID~1565438457) to 5.6~hours.

All observations from PID SCI-20210212-CR-01 were carried out with the intent of being calibrated in full polarisation; to that end, the known polarisation calibrator 3C~286 was observed for two five-minute scans during CBID~1631864177, separated by a broad parallactic angle range. No polarisation calibrators were observed during CBID~1565438457.

Calibration was performed following the same steps as \cite{Riseley2022_MS1455}. To summarise, initial calibration and flagging was carried out using the Containerized Automated Radio Astronomy Calibration (\textsc{caracal}) pipeline\footnote{\url{https://github.com/caracal-pipeline/caracal}} \citep{Jozsa2020,Jozsa2021}. \textsc{caracal} uses the Stimela framework \citep{Makhathini2018} as a wrapper for standard-practice calibration tasks in the Common Astronomy Software Application (\textsc{CASA}) package.

We employed various flagging steps in \textsc{caracal}, including (i) shadowed antennas, (ii) specific channel ranges corresponding to the MeerKAT bandpass edges and known radio frequency interference (RFI) bands, and (iii) automatic sum-threshold flagging with the \verb|tfcrop| algorithm. After initial calibration, we re-flagged our data with \verb|tfcrop| and re-derived calibration tables to refine our solutions, and subsequently applied these to our target.

We then executed an initial round of relatively shallow automated sum-threshold flagging using \textsc{tricolour}\footnote{\url{https://github.com/ska-sa/tricolour}} \citep{Hugo2022}. An initial sky model was generated using \textsc{ddfacet} \citep{Tasse2018} and subtracted from our data; we then re-ran \textsc{tricolour} on the residual data to excise lower-level RFI. Finally, we averaged to a spectral resolution of 1.67~MHz, yielding 512 output channels, and proceeded to self-calibration.

Throughout the self-calibration process, we imaged with \textsc{ddfacet} using \texttt{robust} $=-0.5$ weighting \citep{Briggs1995}, and employed the sub-space deconvolution \citep[\textsc{ssd};][]{Tasse2018} algorithm to improve the modelling of the numerous resolved radio sources across the field of view.

Self-calibration was performed using \textsc{killms} \citep{Tasse2014,Smirnov2015}. We carried out three rounds of phase-only self-calibration and two rounds of amplitude-and-phase self-calibration, both in direction-independent (DI) mode. As in \cite{Riseley2022_MS1455}, we used the quality-based weighting scheme introduced by \cite{Bonnassieux2018} to weight our calibration solutions, which expedited the convergence of our self-cal process. After five rounds of DI self-calibration, our processing had largely converged and we inspected our image products for residual direction-dependent (DD) errors.

To correct for these DD errors, which were more naturally visible in the wider field, we tessellated the sky into 16 regions and carried out two rounds of DD-calibration and imaging, applying amplitude and phase gains on the fly. Finally, we generated an extracted dataset covering a small region around our target by subtracting our best sky model of all sources outside the region of interest. Given that MeerKAT has a large primary beam full-width at half-maximum (FWHM), around 67~arcmin at 1.28~GHz \citep[][]{Mauch2020}, this extraction step allowed us to efficiently post-process our data with manageable overhead.

\subsection{Radio: Giant Metrewave Radio Telescope}
Abell~1413 was observed with the uGMRT on two occasions using the GMRT Wideband Backend \citep[GWB;][]{Reddy2017} with the Band~3 receiver system, which covers the frequency range 250$-$500~MHz. Observations were carried out on 2020~October~03 (project 38\_025; P.I. Cuciti) and 2022~April~22 (project 42\_057; P.I. Biava) for a total on-source time of 7.8~hours. We also observed Abell~1413 using the Band~4 receiver system in the frequency range 550$-$900~MHz on 2022~June~05 (project 42\_057) for a total of 8~hours on-source.

Using recent developments for wide-band data processing, these observations were processed using the Source Peeling and Atmospheric Modelling \citep[\textsc{spam};][]{Intema2009,Intema2017} pipeline. To efficiently process the wide-band data and navigate limitations of the pipeline, the wide-band data are divided into smaller frequency sub-bands and calibrated independently. The pipeline initially derives calibration solutions from the primary calibrators and then applies them to the target field before proceeding to self-calibration. The self-calibration process corrects for both DI and DD calibration errors using a single reference model for all sub-bands, which is obtained by processing the narrow-band GMRT Software Backend (GSB) data recorded alongside the GWB data. The GSB data is processed using the standard \textsc{spam} pipeline. Finally, the calibrated GWB visibilities then ready for postprocessing using other software tools, as described in \S~\ref{sec:postprocessing}.

We note that in this paper we only present high-resolution images from our uGMRT data. The presence of several bright resolved radio galaxies in the wider field limits the dynamic range and image fidelity of our data products, despite our use of the well-verified \textsc{spam} software to perform initial data processing and self-calibration. This effect is particularly pronounced on short baselines, and is worse in Band~4 due to the decreased primary beam FWHM. Techniques for post-processing \textsc{spam}-calibrated datasets are under development but their application is beyond the scope of this paper. We will revisit this dataset in future as part of a wider follow-up campaign of mini-haloes, including the `MeerKAT-meets-LOFAR' sample.

\subsection{Radio: LOFAR HBA}
Abell~1413 was observed with LOFAR as part of LoTSS, and performed using the full International LOFAR Telescope \citep[ILT;][]{vanHaarlem2013} in \texttt{HBA\_DUAL\_INNER} mode, covering the frequency range 120$-$168~MHz. Three LoTSS fields overlap Abell~1413 (P177+22, P178+25, P180+22); however, due to particularly poor ionospheric conditions during observations of P178+25, only the data for P177+22 and P180+22 are used in this work. These two fields are separated from Abell~1413 by an angular distance of 1.68 and 1.95 degrees, respectively. P177+22 was observed on 2017 May 04; P180+22 was observed on 2017 August 09. We note that these are different observations to those presented by \cite{Savini2019}, although this is fundamentally the same set of observations presented by \cite{Lusetti_Tesi2021} and Lusetti et al. (in prep.); we have undertaken an independent postprocessing.

In this work, we only make use of the LoTSS data from the Dutch LOFAR array (Core and Remote stations, encompassing baselines out to $\sim80$~km). The full ILT observations presented in \cite{Riseley2022_MS1455} of MS~1455.0$+$2232 showed that a significant amount of the flux from the radio counterpart to the BCG was lost due to the relatively faint nature of the source; in Abell~1413, the radio galaxies embedded in the mini-halo are significantly fainter than the radio BCG in MS~1455.0$+$2232, and thus we do not expect to make a high signal-to-noise ratio detection with the full ILT.

The Dutch LOFAR array data were processed using the standard LoTSS pipeline\footnote{\url{github.com/mhardcastle/ddf-pipeline/}}, which is described in detail by \cite{Shimwell2019_LOTSS_PaperII,Shimwell2022_LOTSS_PaperIII} and \cite{Tasse2021}. As a brief summary, this pipeline performs flagging, initial calibration, and both DI and DD self-calibration using \textsc{killms} and \textsc{ddfacet}. An extracted dataset was then created, containing a region within 0.35~deg radius around Abell~1413 using the process described by \cite{vanWeeren2021} to allow for efficient re-imaging with different weighting schemes and \textit{uv}-selection ranges.

\subsection{Ancillary Radio Data}
To provide additional flux density measurements for radio sources of interest, we turned to ancillary data. This included a 3~GHz mosaic image from the Karl G. Jansky Very Large Array (VLA) Sky Survey \citep[VLASS;][]{Lacy2020}, sourced via the Canadian Initiative for Radio Astronomy Data Analysis (CIRADA) image cutout server. The radio counterpart to the BCG is undetected in the VLASS mosaic, although we note a marginal detection of a source that may be correspond to the radio core of the embedded head-tail radio galaxy.

In addition, we used the L-band (1008$-$1968~MHz) VLA images of Abell~1413 recently presented by \cite{Osinga2022}. These observations were taken as part of project 15A$-$270, and were performed on 2015~Feb.~02 with the VLA in B configuration, for a total on-source time of 40~minutes. We refer the reader to \cite{Osinga2022} for details of the data processing steps. Note that these observations were not used to study the diffuse emission of the mini-halo, due to the limited sensitivity to extended low surface-brightness emission of the VLA B-configuration observations; instead, these data were used to provide flux density measurements for the various radio galaxies in the vicinity of Abell~1413 --- principally the BCG.

\subsection{Radio Postprocessing}\label{sec:postprocessing}
We followed the same postprocessing steps as \cite{Riseley2022_MS1455}. In brief, we used \textsc{wsclean} \citep{Offringa2014,Offringa2017} version 2.10.0\footnote{WSclean is available at \url{https://gitlab.com/aroffringa/wsclean}} to generate science-quality images from our extracted MeerKAT, LOFAR, and uGMRT datasets. 

We used multi-scale clean in order to optimally model the diffuse emission present in the field. We cleaned using the \texttt{-auto-mask} and \texttt{-auto-threshold} functionality to automate the deconvolution, and used the \texttt{-join-channels} and \texttt{-channels-out} options to improve the wide-band deconvolution, producing Stokes $I$ sub-band images across the bandwidth, in addition to multi-frequency-synthesis (MFS) images at frequencies of  1283~MHz for MeerKAT, 145~MHz for LOFAR, 400~MHz for uGMRT Band 3 and 675~MHz for uGMRT Band 4. 

We also employed a common inner \textit{uv}-cut of $80\uplambda$ for all datasets. This choice reduces contamination from Galactic emission and residual RFI by deselecting baselines between the two substations of each LOFAR HBA Core Station.

Finally, we performed three rounds of additional DI self-calibration on our MeerKAT data using the LOFAR Default Pre-Processing Pipeline \citep[\textsc{dppp};][]{vanDiepen2018}. In each round, we solved for a diagonal gain matrix. Further self-calibration did not yield appreciable improvement. LOFAR data undergo DI self-calibration as part of the extraction process; as such, further self-calibration did not yield appreciable improvement.

Final representative low- and high-resolution images were produced by varying the \texttt{robust} parameter between different \textsc{wsclean} imaging runs. We used \texttt{robust}~$=-0.5$ for low-resolution imaging and \texttt{robust}~$=-2.0$ (corresponding to \texttt{uniform} weighting) for high-resolution imaging. For our uGMRT data, we use only high-resolution images made with \texttt{robust}~$=-2.0$ to study the spectral properties of the nearby compact sources, in particular the embedded radio BCG.

\subsubsection{Flux Scaling}
LOFAR observations and uGMRT observations that are processed using the \textsc{spam} pipeline are tied to the \cite{ScaifeHeald2012} flux scale, which is consistent with the \cite{Kellermann1966} scale above 325~MHz. The MeerKAT data presented in this work had the flux density scale set using observations of PKS~B0407$-$658, and so these data are tied to the \cite{Baars1977} scale. To convert our MeerKAT data to be consistent with the \cite{ScaifeHeald2012} scale, we use a polynomial fit to the values presented in Table~7 of \cite{Baars1977}, performed in log-linear space. This polynomial fit yielded a conversion factor of 0.968 at the reference frequency of our MeerKAT observations ($\nu_{\rm{ref}} = 1283$~MHz).

\begin{figure*}
\begin{center}
\includegraphics[width=0.99\linewidth]{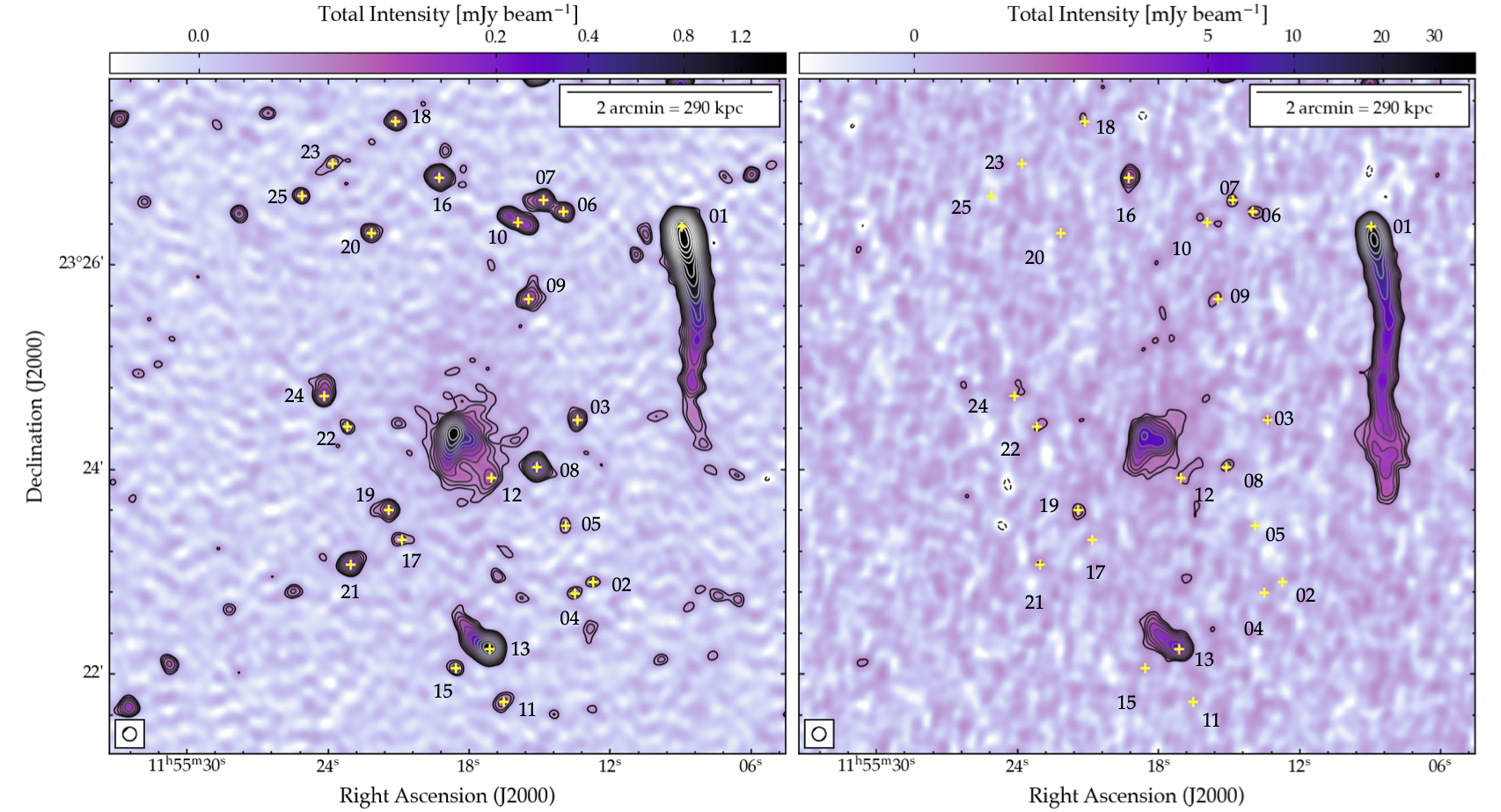}
\cprotect\caption{Radio continuum images of Abell~1413 with MeerKAT at 1283~MHz (\textit{left}, \texttt{robust} $=-0.5$), LOFAR HBA at 145~MHz (\textit{right}, \texttt{robust} $=-1.0$) at 8~arcsec resolution. Contours start at $4\sigma$ and scale by a factor $\sqrt{2}$, where $\sigma = 6.6~\upmu$Jy beam$^{-1}$ at 1283~MHz, and $164~\upmu$Jy beam$^{-1}$ at 145~MHz. The dashed contour denotes the $-3\sigma$ level. Yellow `+' signs identify sources referred to in Table~A1, along with their associated ID numbers, except the embedded head-tail radio galaxy (source 14). All identified sources were subtracted when generating our low-resolution images, with the exception of source 01.}
\label{fig:A1413_continuum}
\end{center}
\end{figure*}

We also used established routines to verify the flux scale of our extracted LOFAR images through comparison with the well-verified LoTSS flux scale. This routine is described in detail elsewhere \citep{Hardcastle2016,Shimwell2019_LOTSS_PaperII,Shimwell2022_LOTSS_PaperIII}, but in brief, we generated an image at 6~arcsec resolution using \textsc{wsclean} and extracted a catalogue using the Python Blob Detection and Source Finder software \citep[\textsc{pybdsf};][]{MohanRafferty2015}, which was compared with a point-source-filtered catalogue derived from the full-field LoTSS image, before performing a linear regression best-fit in the flux:flux plane. Overall, this routine yielded a bootstrapping factor of 1.156 to align with the LoTSS flux scale. Finally, we adopt a typical 5 per cent uncertainty in our MeerKAT flux scale; and a representative 10 per cent uncertainty in our LOFAR flux scale \citep[following][]{Shimwell2022_LOTSS_PaperIII}. We also adopt a 10 per cent systematic uncertainty in our GMRT flux scale.

\subsubsection{Source Subtraction and Final Imaging}
Several compact or partially-resolved sources are visible in the region of the mini-halo in Abell~1413, including the compact radio source associated with the BCG (see Figure~\ref{fig:Abell1413_rgb}) and the nearby head-tail radio galaxy embedded in the mini-halo. Subtraction of these contaminating sources was necessary to fully explore the diffuse emission of the mini-halo.

We followed the same process as \cite{Riseley2022_MS1455} and subtracted the clean component model corresponding to these sources. To generate this model, we imaged with \textsc{wsclean}, applying an inner \textit{uv}-cut of $5{\rm{k}}\uplambda$ to filter the diffuse emission of the mini-halo. This scale corresponds to an angular scale of 41~arcsec or a linear scale of 99~kpc, and was chosen as it effectively suppressed our recovery of the mini-halo without reducing sensitivity to emission from the embedded head-tail radio galaxy (which has a projected largest angular size of 35~arcsec at 1283~MHz).

After subtracting these clean component model of these sources in the \textit{uv}-plane, we generated source-subtracted images using \textsc{wsclean} as per \S\ref{sec:postprocessing}. We produced images at 15~arcsec resolution by using \texttt{robust}~$=-0.5$ in conjunction with appropriate \textit{uv}-tapering and subsequent image-plane smoothing. We summarise the properties of our final images in Table~\ref{tab:img_summary}.

\subsection{X-ray: \textit{Chandra}}
To provide the critical window into the thermal properties of the ICM of Abell~1413, we used X-ray data from \emph{Chandra} using the Advanced CCD Imaging Spectrometer I-array (ACIS-I) instrument. Abell~1413 has been observed with \emph{Chandra} ACIS-I on five occasions (ObsIDs 537, 1661, 5002, 5003, 7696) for a total net exposure time of 128~ks, although some of these ObsIDs are off-axis. These data have been previously presented in \cite{Botteon2018_ChandraProfiles}, and we refer the reader to this paper for full details of the data processing steps; we use the surface brightness and temperature maps originally presented by those authors in this work. Later in this paper we show these maps in Figure~\ref{fig:A1413_boxes} to aid context.

We note that \cite{Botteon2018_ChandraProfiles} performed an edge searching using a Gaussian Gradient Magnitude \citep[GGM;][]{Sanders2016b} filter, which suggested some discontinuities but were not supported by surface brightness profile fits. As with \cite{Riseley2022_MS1455}, we further explored the ICM surface brightness distribution by applying an adaptive GGM filter \citep{Sanders2021} to the \emph{Chandra} mosaic. Unlike our previous investigation of MS~1455.0$+$2232 \citep{Riseley2022_MS1455} however, no statistically-significant edges were found which could indicate the presence of a large-scale sloshing spiral. Therefore we do not present images of our adaptive-GGM filtered map in this paper.

\section{Results}\label{sec:results}

\subsection{Radio continuum properties}
We present our final full-resolution (8~arcsec) maps of Abell~1413 in Figure~\ref{fig:A1413_continuum} at reference frequencies of 1283~MHz (MeerKAT; \textit{left panel}) and 145~MHz (LOFAR; \textit{right panel}). These maps respectively have an rms noise of $6.6~\upmu$Jy~beam$^{-1}$ and $164~\upmu$Jy~beam$^{-1}$.

Through a combination of deeper observations and the application of our advanced data processing recipes, we achieve greater sensitivity compared to previously published studies. Our LOFAR maps at 145~MHz show a factor $\sim2$ improvement in rms at 145~MHz compared to \cite{Savini2019}, who report $270~\upmu$Jy~beam$^{-1}$ at $\sim10$~arcsec resolution. However, we note that this is a factor $\sim2$ worse than the median rms noise of $83~\upmu$Jy~beam$^{-1}$ at 6~arcsec resolution achieved by LoTSS DR2 \citep{Shimwell2022_LOTSS_PaperIII}. This largely reflects the reduction in sensitivity of aperture arrays like LOFAR when observing at low elevation; such an increase in typical rms can also be seen in Fig.~2 of \cite{Shimwell2022_LOTSS_PaperIII} for example. Our MeerKAT 1283~MHz map shows a $\sim70\%$ improvement in rms compared to \cite{Trehaeven2023}, who report $11.2~\upmu$Jy~beam$^{-1}$ at a resolution of $(12.2 \times 5.9)$~(arcsec~$\times$~arcsec); alternatively, our MeerKAT map is around a factor 15 more sensitive than the results reported by \cite{Govoni2009} at a similar frequency.

\begin{figure*}
\begin{center}
\includegraphics[width=0.99\linewidth]{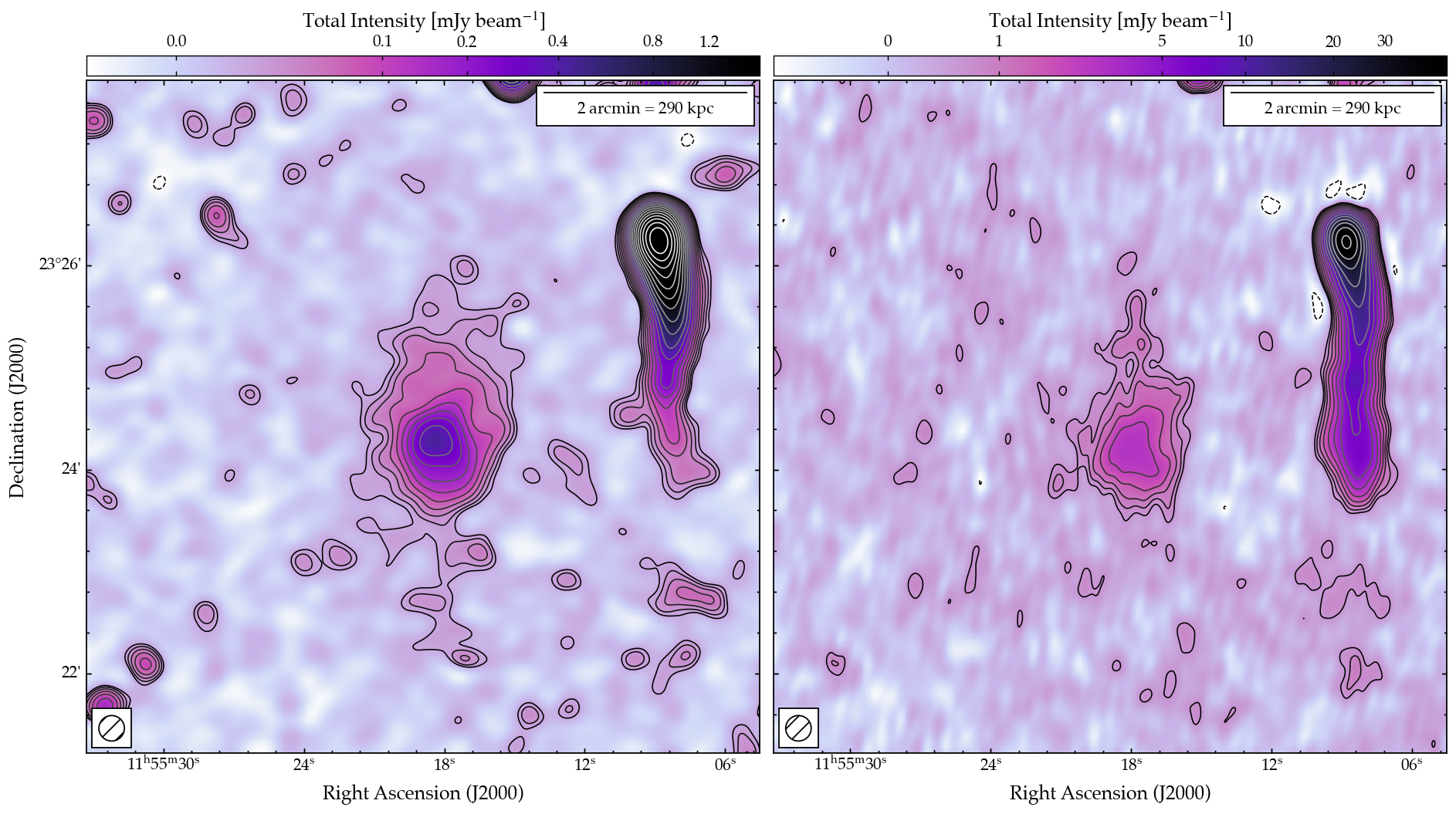}
\cprotect\caption{Source-subtracted radio continuum images of Abell~1413 with MeerKAT at 1283~MHz (\textit{left}) and LOFAR HBA at 145~MHz (\textit{right}) at 15~arcsec resolution. Contours start at $3\sigma$ and scale by a factor $\sqrt{2}$, where the respective value of $\sigma$ is listed in Table~\ref{tab:img_summary}. The dashed contour denotes the $-3\sigma$ level.}
\label{fig:A1413_sub}
\end{center}
\end{figure*}

The high sensitivity of our MeerKAT data enable us to detect many compact radio galaxies in the vicinity of Abell~1413. Beyond these sources, we also recover the moderately-extended head-tail radio galaxy embedded in the mini-halo emission, a second tailed radio galaxy directly to the south of the cluster, as well as the known extended head-tail radio galaxy to the west of the cluster, first reported by \cite{Savini2019}. These sources are clearly visible in Figures~\ref{fig:Abell1413_rgb} and \ref{fig:A1413_continuum}, for example.

Comparatively, our LOFAR map shows fewer sources in the vicinity. Of the compact sources detected by MeerKAT, only a handful are detected at above $4\sigma$. These sources are likely active radio galaxies with typical spectral index values flatter than our `noise spectral index' (essentially the steepest spectral index a source detected at 1283~MHz would have while still remaining detectable at 145~MHz), which is around $-1.5$. Deeper observations with LOFAR would be required to measure their spectral properties.

Overall, the mini-halo is only moderately recovered at 8~arcsec resolution in each map. Some hints of the extended emission are visible in the coherent structure of the local noise, albeit below the $4\sigma$ level, and are also visible in Figure~\ref{fig:Abell1413_rgb}. However, given the extended structure of the embedded head-tail radio galaxy as well as the nearby compact radio sources, these contaminants must be excised before performing analysis of the diffuse mini-halo. 

We also report the properties of these sources, including their reference coordinates, flux density measurements, and optical cross-identifications (where available) in Table~\ref{tab:measurement}. For compact sources, the reference coordinates are given as the best-fit position; for resolved sources, the reference coordinate is the best-fit to the centroid of the brightest component (typically believed to be the radio core).

\begin{table}
\footnotesize
\centering
\caption{Summary of image properties for images of Abell~1413. Images marked with a $^{\dag}$ were produced after source-subtraction, with the application of a \emph{uv}-taper to achieve the desired resolution. The quoted RMS noise values were derived as the average of several off-source regions in the vicinity of the phase centre. \label{tab:img_summary}}
\begin{tabular}{lccccc}
\hline
Telescope    & Freq. & \texttt{Robust} & RMS noise & Resolution & PA \\
  	     & $[$GHz$]$ & &  $[\upmu$Jy beam$^{-1}]$  & $[$arcsec$]$ & $[\degree]$ \\
\hline\hline
\multirow{1}{*}{VLA} & 1.512 & $0.0$ & 18.2  & $4.0\times3.1$ & 34 \\
\hline
\multirow{3}{*}{MeerKAT} & \multirow{3}{*}{1.283} & $-0.5$ & 6.6  & 8  & 0 \\
            & & $-2.0$ & 23.7 & $6.7\times2.9$ & 178 \\
            & & $-0.5^{\dag}$ & 6.9  & 15 & 0 \\
\hline
\multirow{2}{*}{uGMRT} & 0.675 & $-2.0$ & 17.8  & $3.5\times2.5$ & 64 \\
                        & 0.400 & $-2.0$ & 47.1  & $6.1\times3.0$ & 54 \\
\hline
\multirow{3}{*}{LOFAR} & \multirow{3}{*}{0.145} & $-0.5$  & 163  & 8 & 0 \\
            & & $-2.0$  & 566  & $3.9\times2.6$ & 102 \\
            & & $-0.5^{\dag}$  & 192  & 15 & 0 \\
\hline
\end{tabular}
\end{table}


\subsection{Source-subtracted images}
Figure~\ref{fig:A1413_sub} presents our images of Abell~1413 at 15~arcsec resolution, after subtracting the contaminating compact and marginally-resolved radio galaxies identified in Figure~\ref{fig:A1413_continuum}. These images were produced using a combination of $uv$-tapering and image-plane convolution to achieve this resolution. At 15~arcsec resolution, we measure an rms noise of $6.9~\upmu$Jy~beam$^{-1}$ with MeerKAT at 1283~MHz and $192~\upmu$Jy~beam$^{-1}$ with LOFAR at 145~MHz.

The enhanced sensitivity to diffuse emission provided by these techniques allows us to detect the extended mini-halo with high significance in Figure~\ref{fig:A1413_sub}. The mini-halo is extended and highly asymmetrical, being elongated along a north-south axis, similar to the disturbed BCG, and following the same axis as both the galaxy substructure \citep{Castagne2012} and the X-ray surface brightness distribution. This has also been commented on by previous radio studies \citep{Savini2019,Lusetti_Tesi2021,Trehaeven2023}.

\begin{figure*}
\begin{center}
\includegraphics[width=0.8\linewidth]{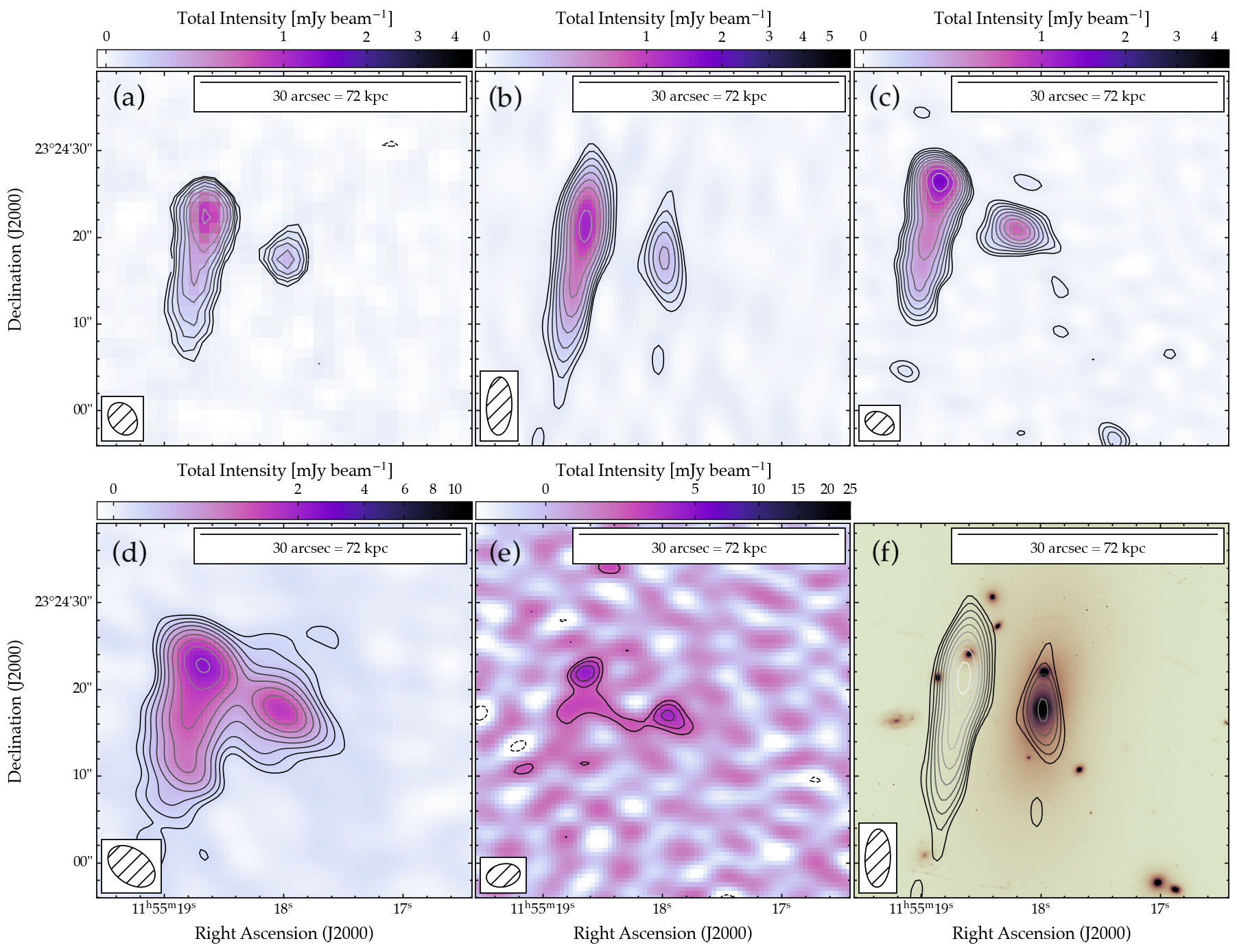}
\cprotect\caption{Zoom on the BCG of Abell~1413. Panels (a) through (e) show radio data as both colourmaps and contours, whereas panel (f) shows an \emph{HST} WFPC2 image, combining data from the $F850LP$ and $F775W$ filters overlaid with MeerKAT radio data from panel (b). The radio data shown are as follows: VLA data at 1519~MHz (panel a), MeerKAT data at 1283~MHz (panel b), uGMRT Band 4 data at 675~MHz (panel c), uGMRT Band 3 data at 400 MHz (panel d) and LOFAR data at 145~MHz (panel e). All radio maps were produced using \texttt{robust} $= -2.0$ weighting, except in panel (a), where \texttt{robust} $= 0$ weighting was used. All contours start at $3\sigma$ increment by a factor of $\sqrt{2}$, where $\sigma$ is listed in Table~\ref{tab:img_summary}.}
\label{fig:A1413_BCG}
\end{center}
\end{figure*}

In both our MeerKAT and LOFAR source-subtracted maps, it is curious to note that the mini-halo appears to be divided into two slightly distinct regions. The inner region of the mini-halo, close to the BCG\footnote{Within a radius of 21~arcsec, or 50.7~kpc given our cosmology.}, appears brighter and marginally more regular, whereas the larger-scale diffuse emission extends further and is clearly fainter and more diffuse. This may hint at the presence of multiple components, similar to a handful of other cases reported in recent years from studies of relaxed clusters using next-generation interferometers \citep[e.g.][]{Savini2018,Savini2019,Biava2021_RXCJ1720,Riseley2022_MS1455}.

The total extent of contiguous emission above the $3\sigma$ level in our MeerKAT map provides a largest angular size (LAS) of 242~arcsec, equivalent to a largest linear size (LLS) of 584~kpc at the cluster redshift, given our cosmology. This overlaps with a region of faint emission to the south where we cannot exclude the possibility of some residuals associated with the southern tailed radio galaxy; excluding this region we measure a LAS of 186~arcsec (LLS of 449~kpc). Our LLS measurements indicate that the mini-halo in Abell~1413 is at least twice as large as previously measured by \cite{Govoni2009}, who reported a angular size of 90~arcsec (around 220~kpc).

Measuring similarly in our LOFAR map, we recover a LAS around 142~arcsec, equivalent to 343~kpc. Similarly to our previous work on MS~1455.0$+$2232, the apparent decrease in size at 145~MHz is likely related to the relative sensitivity of our MeerKAT and LOFAR maps. The relative sensitivity places a lower-limit on the spectral index of $-1.53$ in regions where MeerKAT measures emission from the mini-halo but LOFAR does not; this is not in tension with the expected spectral index of the mini-halo based on recent detailed studies with high-quality data \citep[e.g.][]{Biava2021_RXCJ1720,Riseley2022_MS1455}.

We note that while the extent of the diffuse emission measured from our LOFAR maps is smaller than the $\sim 800$~kpc extent reported by \citep[][Lusetti et al., in prep.]{Lusetti_Tesi2021} using fundamentally the same LOFAR observations, the results are not in tension. Those authors present a lower-resolution study focused on mapping the extent of the `mini'-halo, hence adopting a lower threshold of $2\sigma$ and more naturally-weighted imaging parameters. Our study is focused on understanding the nature of the particle acceleration mechanism via the spectral properties, and hence we adopt a higher threshold of $3\sigma$ and more robust-weighted imaging parameters to allow for a more spatially-resolved spectral study. When common imaging settings are used, the results are consistent.


\subsection{The brightest cluster galaxy in Abell~1413}
Figure~\ref{fig:A1413_BCG} shows a zoom on the central region of Abell~1413, where the BCG and companion head-tail radio galaxy are visible. We overlay radio contours from our high-resolution maps. The BCG itself, catalogued in the literature as MCG+04-28-097 \citep{Noonan1972}, is a large cD galaxy with extremely high ellipticity \citep{Castagne2012} at redshift $z = 0.1429$ \citep{Humason1956}.

The BCG is detected at high significance in our MeerKAT, JVLA and uGMRT maps in addition to our LOFAR map, where the signal-to-noise ratio is lower due to the reduced sensitivity of LOFAR when adopting \texttt{uniform} weighting. The embedded head-tail radio galaxy is also detected at high significance in our JVLA, MeerKAT, and uGMRT maps; only the presumed core of this head-tail is detected by LOFAR in Figure~\ref{fig:A1413_BCG}.

\begin{figure}
\begin{center}
\includegraphics[width=0.99\linewidth]{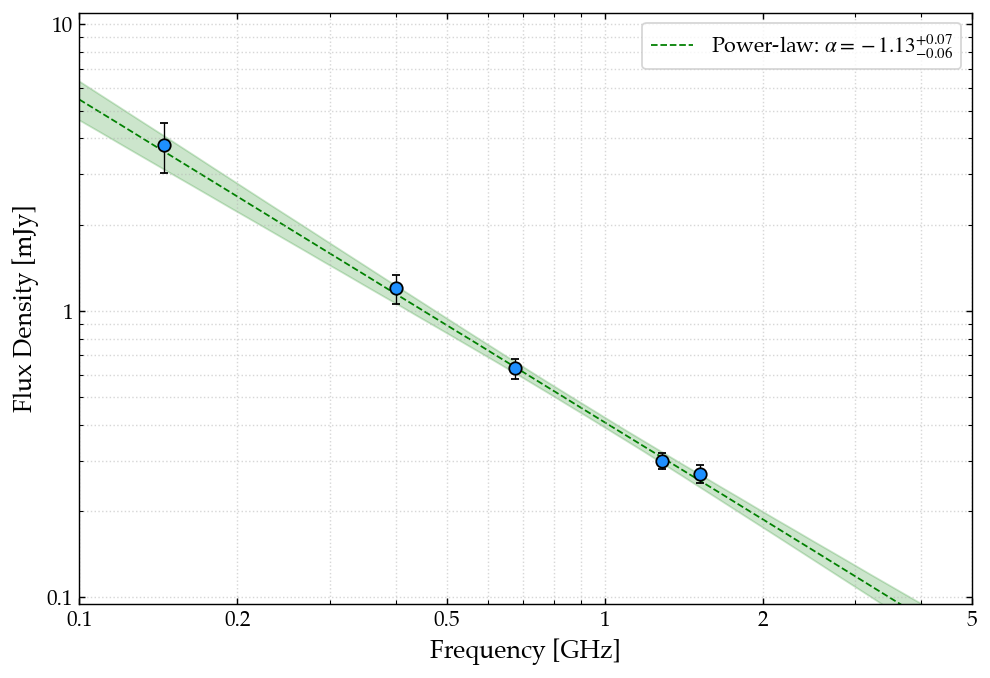}
\cprotect\caption{Integrated spectral energy distribution for the unresolved radio BCG in Abell~1413. Blue datapoints indicate new measurements made on our data. Dashed line and shaded region respectively represent the best-fit power-law spectral index and the uncertainty region corresponding to the 16th and 84th percentiles: $\alpha = -1.13^{+0.07}_{-0.06}$.}
\label{fig:A1413_BCG_SED}
\end{center}
\end{figure}

We used these to measure the flux density of the (unresolved) radio counterpart to the BCG. These measurements are listed in Table~\ref{tab:bcg_fluxes} and presented in Figure~\ref{fig:A1413_BCG_SED}; they indicate no departure from a single power-law behaviour between 145~MHz and 1.519~GHz. We fitted a single power-law model to the data, from which we derive a spectral index $\alpha = -1.13^{+0.07}_{-0.06}$.

This spectral index is steeper than that measured for `typical' active radio galaxies outside of cluster environments, which generally show a canonical synchrotron spectral index of around $-0.8$. It is also significantly steeper than that measured for the BCG of MS~1455.0$+$2232, which was shown to exhibit a spectral break in our first MeerKAT-meets-LOFAR paper \citep{Riseley2022_MS1455}. There the spectral index below the break was very flat, $\alpha_{\rm low} = -0.45 \pm 0.05$, and above the spectral break more typical of `standard' field radio galaxies $\alpha_{\rm low} = -0.81 \pm 0.18$. 

However, some radio galaxies in cluster environments show evidence of similarly steep spectra \citep[e.g.][]{Riseley2022_A3266} although this is more typical of emission in the extended lobes rather than the active cores. In their survey of the radio properties of BCGs, \cite{Hogan2015} find a representative spectral index of $-1.0$ for the `non-core' component of BCG radio emission, which comprises `all other emission' besides the core, largely constituting lobe emission and past AGN activity. As such, given the steep spectrum of $\alpha = -1.13^{+0.07}_{-0.06}$ we measure for the radio counterpart to the BCG in Abell~1413, we consider it more likely that this radio source is lobe-dominated in comparison to the radio counterpart to the BCG in MS~1455.0$+$2232, which is likely core-dominated.

\begin{table}
\centering
\caption{Flux density measurements for the unresolved radio counterpart to the BCG in Abell~1413. All measurements quoted on the \protect\cite{ScaifeHeald2012} flux density scale.
\label{tab:bcg_fluxes}}
\begin{tabular}{lr}
\hline
Frequency    & Flux density     \\
$[$GHz$]$    & $[$mJy$]$        \\
\hline\hline
1.519        & $0.27 \pm 0.02$ \\
1.283        & $0.30 \pm 0.02$ \\
0.675        & $0.63 \pm 0.05$ \\
0.400        & $1.20 \pm 0.14$ \\
0.145        & $3.78 \pm 0.75$ \\
\hline
\end{tabular}
\end{table}

The $k$-corrected radio power $P_{\nu}$ at frequency $\nu$ is expressed as:
\begin{equation}\label{eq:radio_lum}
    P_{\nu} = 4 \pi \,  D_{\rm L}^2 \, S_{\nu} \, (1 + z)^{-(1 + \alpha)}
\end{equation}
where $D_{\rm L}$ is the luminosity distance to the object and $S_{\nu}$ is the flux density at frequency $\nu$. In the case of the BCG in Abell~1413, the redshift $z = 0.1429$ implies $D_{\rm L} = 650.7$~Mpc given our cosmology. Thus, Equation~\ref{eq:radio_lum} yields a 1.4~GHz radio luminosity of $P_{\rm 1.4~GHz} = (1.46 \pm 0.07) \times 10^{22}$~W~Hz$^{-1}$ for the radio counterpart to the BCG of Abell~1413.

\section{Analysis: The Mini-Halo in Abell~1413}\label{sec:analysis}

\subsection{Spectral properties}

\subsubsection{Integrated spectrum}
Measuring the total integrated flux density for the mini-halo in Abell~1413 is non-trivial due to the likely imperfect subtraction of the head-tail radio galaxy to the south of the cluster. We defined the extent of the mini-halo to be bounded by the contiguous $3\sigma$ contour, as measured by MeerKAT and presented in the left panel of Figure~\ref{fig:A1413_sub}, but excluding the southern extremity.

Integrating over this region, MeerKAT recovers a total flux density of $S_{\rm 1283~MHz} = 3.23 \pm 0.17$~mJy. For LOFAR, we measure a total integrated flux density of $S_{\rm 145~MHz} = 29.5 \pm 3.4$~mJy. Thus, our integrated spectral index for the mini-halo in Abell~1413 is $\alpha_{\rm 145~MHz}^{\rm 1283~MHz} = -1.01 \pm 0.06$.

This is somewhat flatter than the estimated spectral index of $\alpha^{\rm 1400~MHz}_{\rm 144~MHz} \sim -1.3$ reported by \cite{Savini2019}. This discrepany is unsurprising, as Savini et al. do not provide an uncertainty estimate, and additionally the 1.4~GHz measurement used by those authors was the VLA flux density measurement from \cite{Govoni2009}, whereas our MeerKAT maps are demonstrably more sensitive. \cite{Trehaeven2023} report a MeerKAT in-band spectral index of $\alpha^{\rm 1.5~GHz}_{\rm 1~GHz} = -1.52 \pm 0.46$ for the mini-halo in Abell~1413, which is broadly consistent with our integrated spectrum, although their uncertainty is large. 

Looking beyond this cluster at the broader population, this is consistent with the spectral index measured for MS~1455.0$+$2232 \citep[$-0.97 \pm 0.05$;][]{Riseley2022_MS1455} and similar to several other clusters studied with the latest generation of radio interferometers \citep[e.g.][]{Raja2020,Timmerman2021,Biava2021_RXCJ1720}, which enable studies to be performed with improved surface brightness sensitivity, resolution, and dynamic range.

\subsubsection{Luminosity and Scaling Relations}
Using the integrated flux density measurements and spectral index derived in the previous section, and taking the luminosity distance at the cluster redshift $D_{\rm L} = 651.2$~Mpc, Equation~\ref{eq:radio_lum} yields a 1.4~GHz radio power of $P_{\rm 1.4~GHz} = (1.50 \pm 0.08) \times 10^{23}$~W~Hz$^{-1}$ for the mini-halo.

\begin{figure}
\begin{center}
\includegraphics[width=0.99\linewidth]{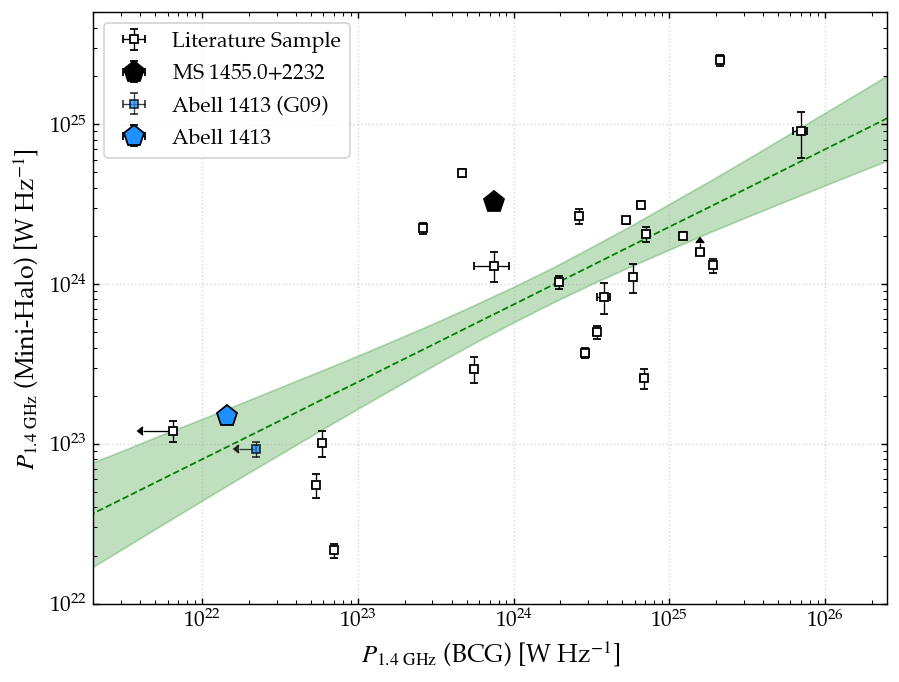}
\cprotect\caption{Scaling plane between BCG radio power and mini-halo radio power at 1.4~GHz. Our new measurements for Abell~1413 are shown by the blue pentagon; the blue square indicates the previous measurements and limits on the mini-halo and BCG radio power, respectively, from \protect\cite{Govoni2009}. The measurements for MS~1455.0$+$2232 are from \protect\cite{Riseley2022_MS1455}. The `Literature Sample' comprises the sample of \protect\cite{Giacintucci2019} plus the mini-halo candidate reported by \protect\cite{Norris2021_EMUPilot}. Uncertainties in the radio power for Abell~1413 and MS~1455.0$+$2232 are not visible at this scale. The dashed line denotes the best-fit scaling relation, which exhibits a slope of $m = 0.48^{+0.11}_{-0.11}$; the shaded region indicates the $1\sigma$ uncertainty on the slope.}
\label{fig:power_scaling}
\end{center}
\end{figure}

To compare our radio power measurement with the known population, we used the most recent compilation of mini-haloes from \cite{Giacintucci2019}, updated with our measurement for MS~1455.0$+$2232 \citep{Riseley2022_MS1455} and the mini-halo candidate reported in the Evolutionary Map of the Universe \citep[EMU;][]{Norris2011} Pilot Survey region \citep{Norris2021_EMUPilot}. We show the known population in the power scaling plane between BCG radio power and mini-halo radio power at 1.4~GHz in Figure~\ref{fig:power_scaling}. This power scaling plane allows us to explore the relation between mini-haloes and AGN feedback processes driven by the BCG. In a scenario where the BCG plays some role in powering mini-haloes, such as by driving turbulence through mechanical feedback and/or by seeding cosmic rays in the ICM, we would expect to observe a correlation. This was explored for the first time in detail by \cite{RichardLaferriere2020}, who found a moderate-strength correlation in this plane.

For context, we also show the position of the mini-halo in Abell~1413 using the measurements reported by \cite{Govoni2009}. We note that these authors did not detect the radio counterpart to the BCG at 1.4~GHz but as such reported an upper limit to the BCG radio power.

We opted to use \textsc{Linmix}\footnote{Currently available at \url{https://linmix.readthedocs.io/en/latest/src/linmix.html}.} \citep{Kelly2007} to fit a scaling relation for the BCG and MH radio power, both at at 1.4~GHz, as shown in Figure~\ref{fig:power_scaling}. Scaling relations for similarly statistically-significant MH populations have been derived in relatively recent years by \cite{RichardLaferriere2020}, although using differing methods: the perhaps more `classic' BCES-orthogonal and BCES-bisector. However, those authors fit the relation using two different abscissas: the `BCG steep' radio power at 1~GHz and the `BCG core' radio power at 10~GHz. In our census we do not have the data to distinguish clearly between these components and instead derive our scaling relation using the total BCG radio power measured at 1.4~GHz from our broad-band data. We derived this using a power-law relation in log-log space of the form:
\begin{equation}
    {\rm log}\left( P_{\rm 1.4~GHz} ({\rm MH}) \right) = a + m \, {\rm log}\left( P_{\rm 1.4~GHz} ({\rm BCG}) \right)
\end{equation}
finding a best-fit slope of $m = 0.48^{+0.11}_{-0.11}$. We find that these two quantities exhibit a moderate-to-strong correlation, as we find Spearman and Pearson coefficients of $r_{\rm S} = +0.56$ and $r_{\rm P} = +0.71$ respectively. We find a stronger correlation than previously reported by \cite{Giacintucci2019}, for example, who found $r_{\rm S} = +0.43$. This increase in correlation strength is likely dominated by the improved sample size considered, as our results show consistency with the correlation strength reported by \cite{RichardLaferriere2020}: $r_{\rm S} = +0.53$ and $r_{\rm P} = +0.68$ for the scaling between MH power at 1.4~GHz and BCG-steep radio power at 1~GHz.

From Figure~\ref{fig:power_scaling} we can see that our new measurements place Abell~1413 toward the faint end of the power scaling plane. The position is not dramatically shifted with respect to previous measurements from \cite{Govoni2009}, and thus Abell~1413 remains among the faintest mini-haloes in the known population.

\subsubsection{Resolved spectral properties}\label{sec:alfa_maps}
\begin{figure*}
\begin{center}
\includegraphics[width=0.95\linewidth]{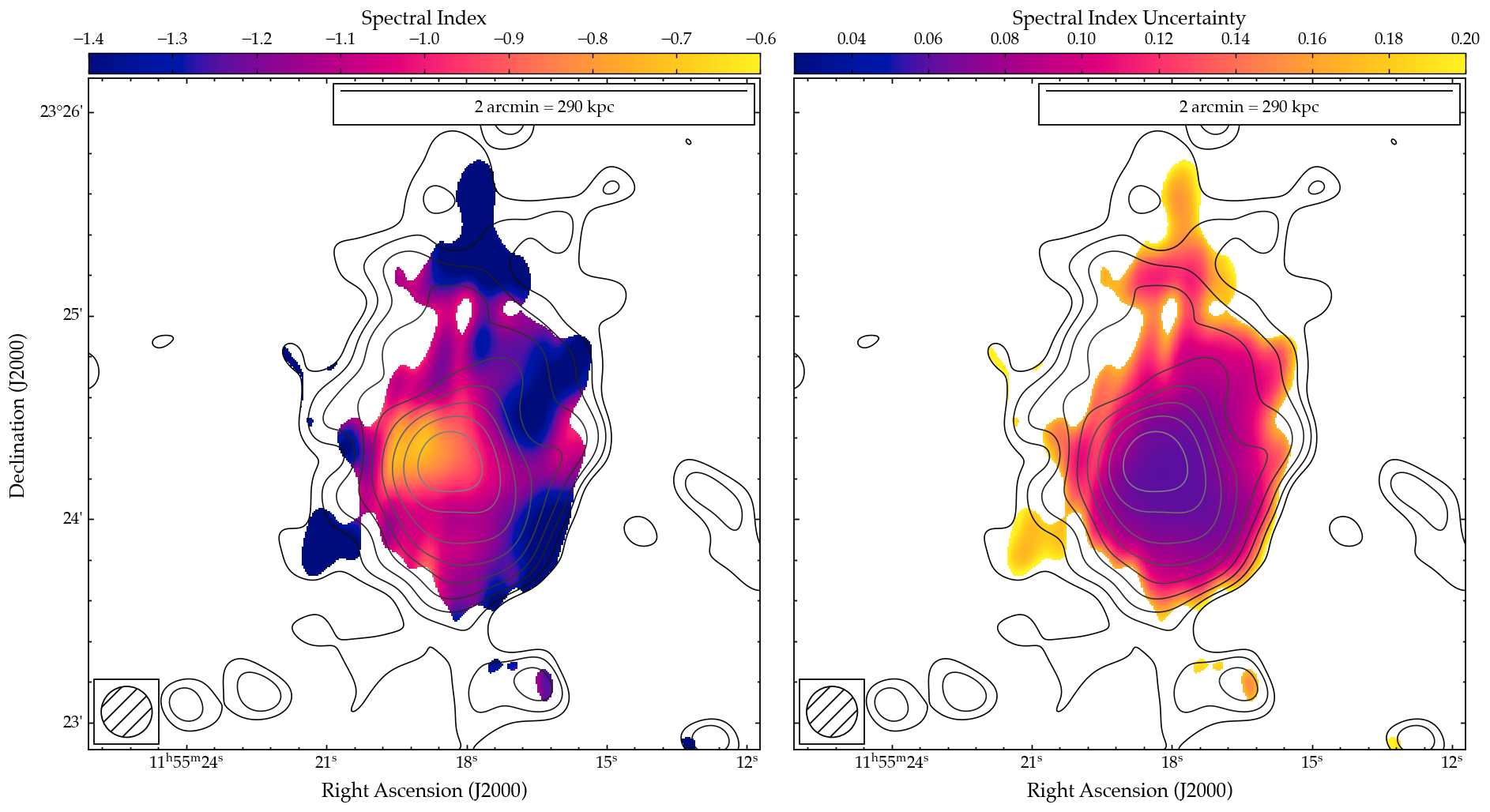}
\cprotect\caption{Spectral index (\textit{left}) and associated uncertainty (\textit{right}) of Abell~1413 between 145 and 1283~MHz at 15~arcsec resolution, made from our source-subtracted maps presented in Figure~\ref{fig:A1413_sub}. Contours denote the source-subtracted MeerKAT data at 1283~MHz, starting at $3\sigma$ and scaling by a factor $\sqrt{2}$.}
\label{fig:A1413_alfa}
\end{center}
\end{figure*}

In Figure~\ref{fig:A1413_alfa} we report the spatially-resolved spectral index map between 1283~MHz and 145~MHz at 15~arcsec resolution, along with the associated uncertainty. While detailed investigation of the entire mini-halo volume is somewhat limited by the reduced sensitivity of our LOFAR data relative to our MeerKAT data, we can still study the resolved spectral properties of much of the mini-halo.

From Figure~\ref{fig:A1413_alfa}, when measuring across the entire mini-halo we find an overall median spectral index of $\langle \alpha \rangle = -1.18 \pm 0.11$, consistent with our integrated spectral index derived in Section~4.1.1. However, this does not tell the whole story, as the spectral index map appears to show two distinct trends. 

The inner region of the mini-halo close to the BCG (within 21~arcsec, or 50.7~kpc) which appears brighter and more uniform in surface brightness, seems to show a distinct trend separate from the more extended diffuse emission. This inner region appears to show a generally flatter spectral index around $-0.7$ to $-1$ whereas the larger-scale diffuse emission appears to show a typical spectral index around $-1.1$ or steeper. 

Measuring the median spectral index both inside and outside this `brighter region' we find $\langle \alpha_{\rm in} \rangle = -0.97 \pm 0.07$ and $\langle \alpha_{\rm out} \rangle = -1.12 \pm 0.13$. While these values are barely consistent at the $\sim 1 \sigma$ level, it suggests tentative evidence of radial steepening in the spectrum of the MH. Additionally, the significance is likely greater, as this uncertainty estimate includes the systematic uncertainty; accounting only for the \textit{statistical} uncertainty (as flux scale errors would not affect the \textit{relative} spectral index) would likely increase the significance.

The lower-limit to the spectral index in these outer regions derived based on the relative sensitivity of our MeerKAT and LOFAR data does not provide powerful constraints in the outermost regions, as these regions must have $\alpha \gtrsim -1.53$ given that they are detected by MeerKAT but not LOFAR. Deeper low-frequency observations would be required to examine this further as we are currently limited by the quality of the available LOFAR HBA data. However, this does suggest that there is no ultra-steep spectrum highly-diffuse component as observed in some other cases \citep[e.g.][]{Biava2021_RXCJ1720}.

\begin{figure}
\begin{center}
\includegraphics[width=0.95\linewidth]{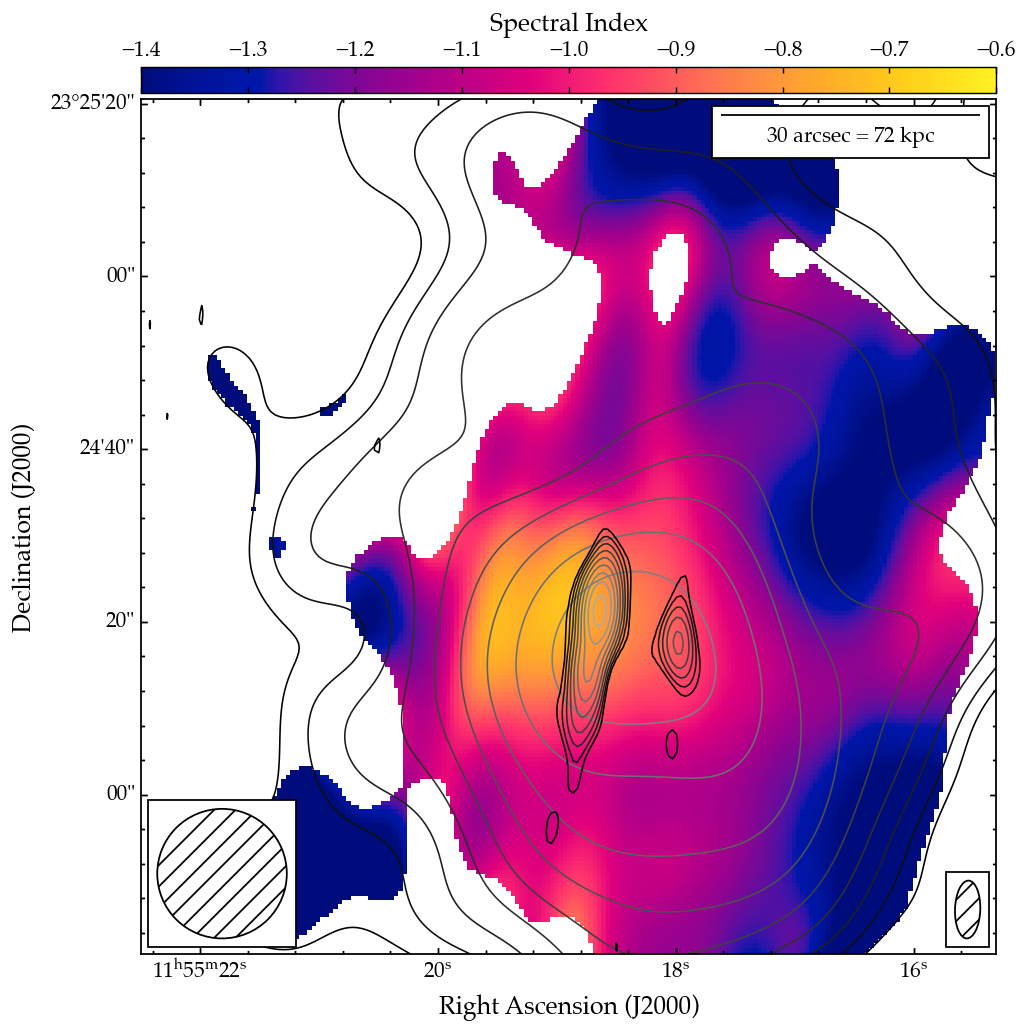}
\cprotect\caption{Zoom on the spectral index map of Figure~\ref{fig:A1413_alfa}, showing the central region around the embedded radio galaxies. Contours show the low-resolution source-subtracted MeerKAT surface brightness at 15~arcsec resolution as well as the high-resolution MeerKAT surface brightness from our \texttt{robust -2} weighted maps, showing the embedded radio sources. The beam sizes of each map are indicated in the lower-left and lower-right corners, respectively.}
\label{fig:A1413_alfa_zoom}
\end{center}
\end{figure}

Figure~\ref{fig:A1413_alfa_zoom} presents a zoom on the spectral index map in the left-hand panel of Figure~\ref{fig:A1413_alfa}, with our high-resolution MeerKAT map (Figure~\ref{fig:A1413_BCG}, panel b) overlaid as contours. From this Figure, we see that the region of flatter spectrum within the MH is offset to the East of the embedded radio sources, in particular the head-tail radio galaxy. As such, we consider it unlikely that this region indicates the presence of residual emission from an inadequate source model, but rather represents a flatter-spectrum component of diffuse emission within the MH.

\begin{figure*}
\begin{center}
\includegraphics[width=0.99\linewidth]{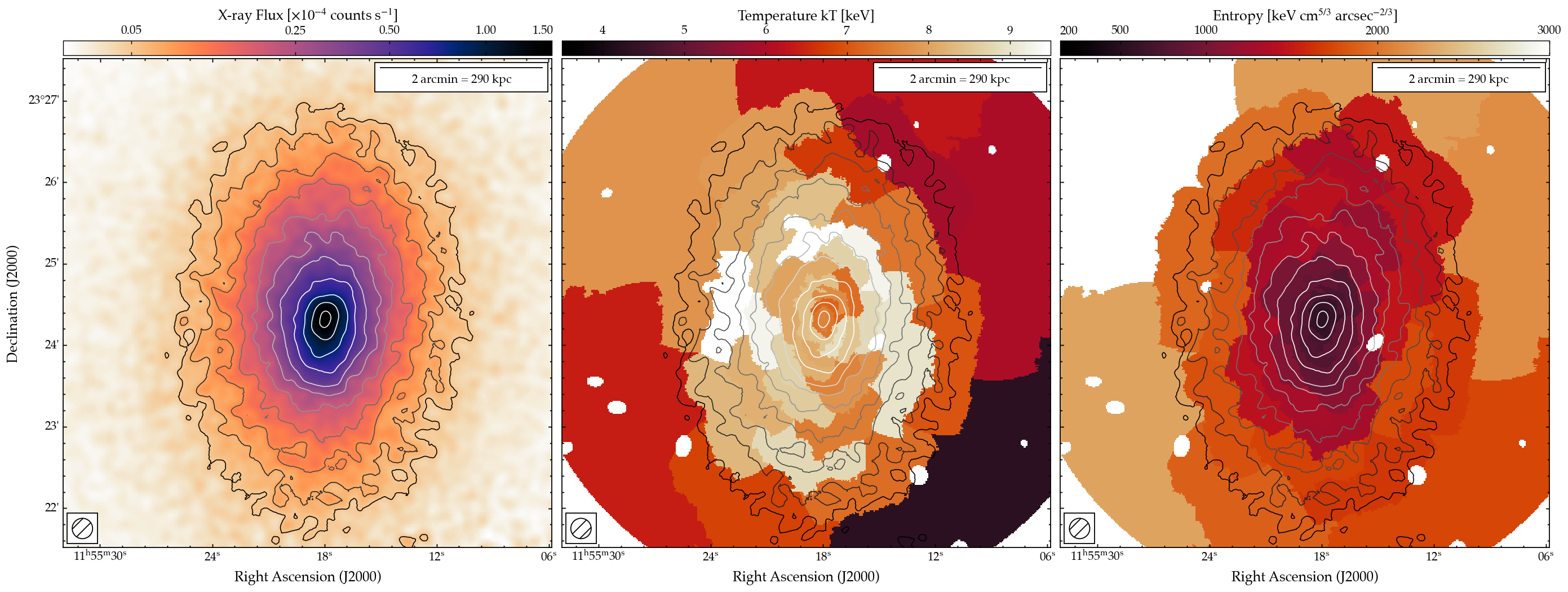}
\includegraphics[width=0.75\linewidth]{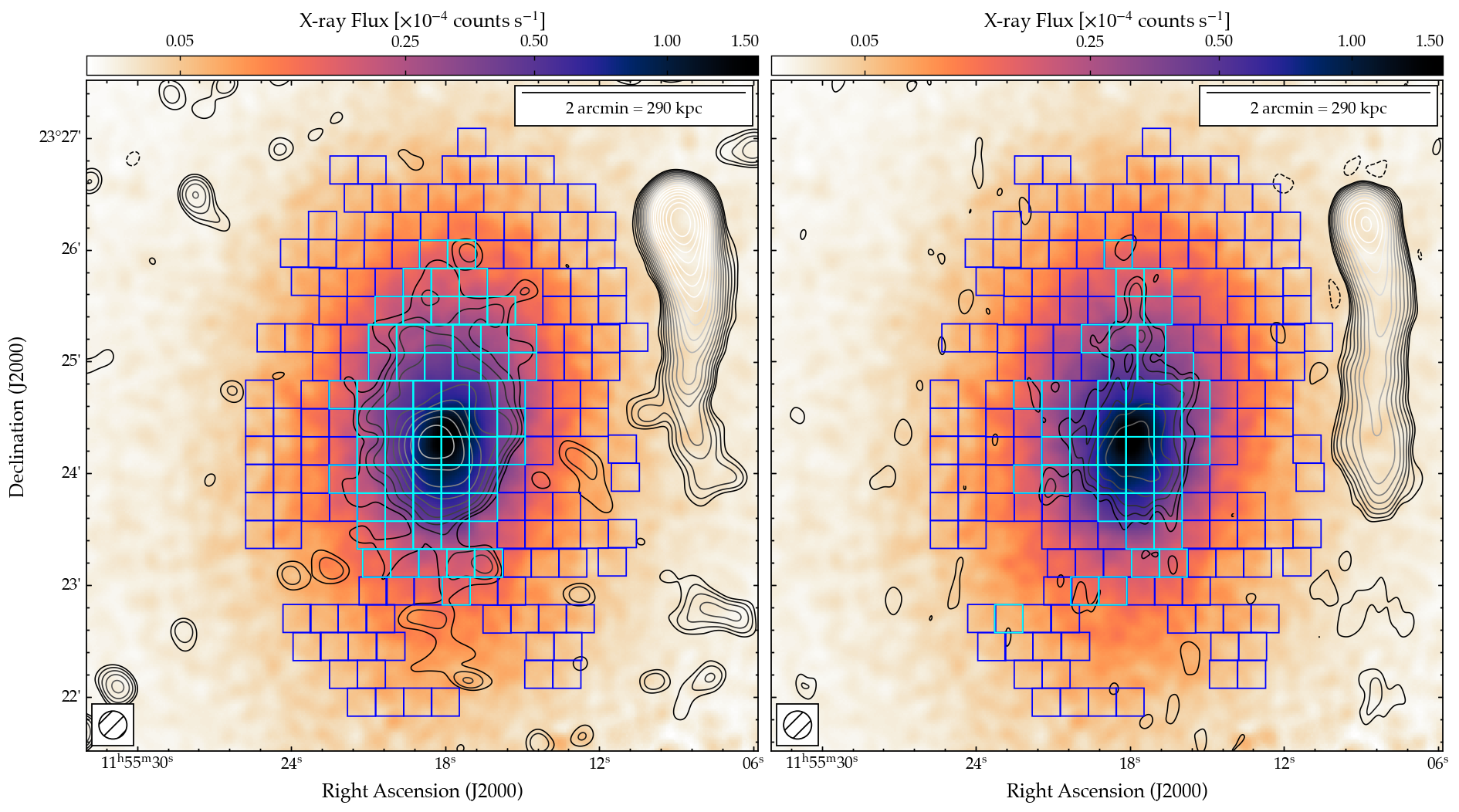}
\cprotect\caption{Radio/X-ray overlay images of Abell~1413. \textit{Top row:} X-ray surface brightness in the 0.5$-$2~keV band measured by \textit{Chandra}, smoothed with a 15~arcsec FWHM Gaussian (\textit{left}); ICM pseudo-temperature (\textit{centre}); pseudo-entropy (\textit{right}). These maps were derived and presented in Fig.15 of \protect\cite{Botteon2018_ChandraProfiles} and are shown here to aid context. Contours show the X-ray surface brightness starting at $5\times10^{-6}$~counts~s$^{-1}$ and scaling by a factor of $\sqrt{2}$. With a central temperature of $kT \simeq 7.5$\,keV and central entropy of $K \simeq 564$\,keV\,cm$^{5/3}$\,arcsec$^{-2/3}$, Abell~1413 exhibits characteristics that are neither typical of fully relaxed or merging clusters. \textit{Bottom row:} The colour map shows the X-ray surface brightness as per the left panel in the upper row. Contours denote source-subtracted radio data at 15~arcsec resolution as per Figure~\ref{fig:A1413_sub} (\textit{left}: MeerKAT at 1283~MHz; \textit{right}: LOFAR at 145~MHz). Boxes show the 15 arcsec square regions used to profile the radio/X-ray correlations: cyan and blue boxes show regions where the radio surface brightness is above and below $3\sigma$ level, respectively.}
\label{fig:A1413_boxes}
\end{center}
\end{figure*}

\subsection{Thermal/non-thermal comparison}
Observationally, both radio haloes and mini-haloes follow a similar morphology to the X-ray ICM. Given this connection, we expect a correlation between the observational properties of the \emph{non-thermal} components --- the CRe and magnetic field, traced by the diffuse synchrotron emission --- and the \emph{thermal} components --- the hot plasma of the ICM, traced by the bremsstrahlung X-ray emission.

Figure~\ref{fig:A1413_boxes} presents a radio/X-ray overlay of Abell~1413. In the top row we show the X-ray surface brightness measured by \textit{Chandra} as well as the X-ray pseudo-temperature (kT) and pseudo-entropy (K) maps originally presented by \cite{Botteon2018_ChandraProfiles} to provide context for the reader. In the bottom row we show the X-ray surface brightness with our source-subtracted radio contours overlaid. As expected, the mini-halo fills much of the volume of the X-ray emitting ICM. To quantitatively explore the thermal/non-thermal connection, we placed adjacent 15~arcsec square boxes across the extent of the X-ray emission recovered by \chandra{}, above a level of $5\times10^{-6}$~counts~s$^{-1}$. However, to avoid contamination from imperfect subtraction of the more complex sources in the vicinity of the mini-halo, we excised those regions where residuals above the $3\sigma$ level remained. Our final region set is also shown in Figure~\ref{fig:A1413_boxes}. Following previous works, we adopted a $2\sigma$ level as the threshold between measurements and limits; regions where the average radio surface brightness was above the $2\sigma$ level we took as measurements (cyan boxes in Figure~\ref{fig:A1413_boxes}), regions where the average radio surface brightness was below this level we took as $2\sigma$ limits (blue boxes in Figure~\ref{fig:A1413_boxes}). In each case, we use the median value of the appropriate quantities in each region to examine the correlations; in each case, changes in the placement of the regions have no significant effect on the fit result.

\subsubsection{Point-to-point correlation: surface brightness}
\begin{figure*}
\begin{center}
\includegraphics[width=0.95\linewidth]{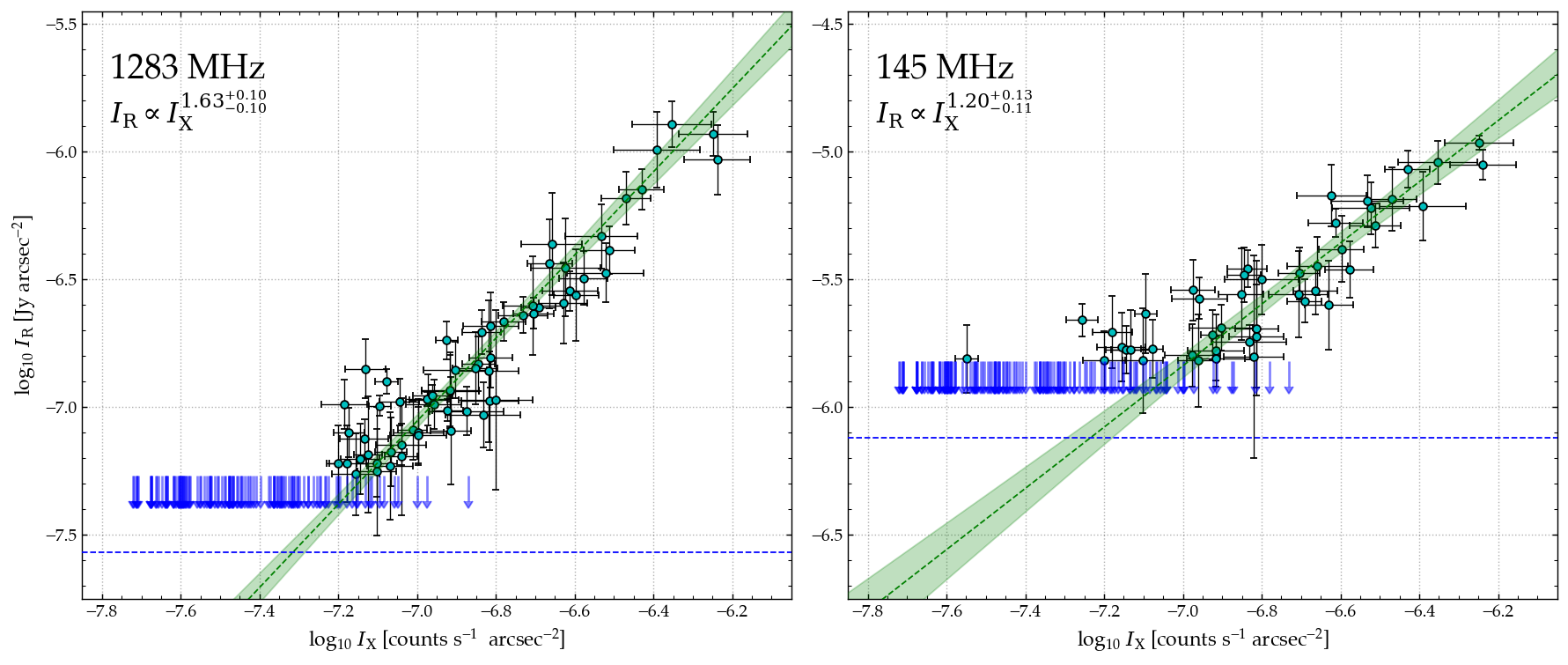}
\cprotect\caption{Radio/X-ray surface brightness correlation ($I_{\rm{R}}/I_{\rm{X}}$) for the mini-halo in Abell~1413 at 1283~MHz (\textit{left panel}) and 145~MHz (\textit{right panel}) at 15~arcsec resolution. Datapoint markers are colourised according to the extraction region, with blue arrows denoting the $2\sigma$ upper limit adopted for regions where the average surface brightness is below $3\sigma$. The dashed blue indicates the $1\sigma$ level. Dashed green line shows the best-fit power-law relation (Equation~\ref{eq:ptp_correlation}) derived using \textsc{Linmix}, with the $1\sigma$ uncertainty region shown in shaded green. The slope of the best-fit power-law is $b_{\rm{1283~MHz}} = 1.63^{+0.10}_{-0.10}$ at 1283~MHz and $b_{\rm{145~MHz}} = 1.20^{+0.13}_{-0.11}$ at 145~MHz.}
\label{fig:ptp_radio_xray}
\end{center}
\end{figure*}

Figure~\ref{fig:ptp_radio_xray} presents the radio/X-ray surface brightness correlation --- the $I_{\rm{R}}/I_{\rm{X}}$ plane --- for Abell~1413, as measured from the regions shown in Figure~\ref{fig:A1413_boxes}. Our data appear to show a strong and positive correlation between $I_{\rm{R}}$ and $I_{\rm{X}}$, reflecting the nature of the correlation between the non-thermal and thermal components. 

Indeed, we find Spearman and Pearson coefficients of $r_{\rm S} = +0.90$ and $r_{\rm P} = +0.94$ at 1283~MHz and $r_{\rm S} = +0.84$ and $r_{\rm P} = +0.86$ at 145~MHz, indicating a very strong correlation. To quantify the slope of the correlation, we fit a power-law relation (in log-log space) of the form:
\begin{equation}\label{eq:ptp_correlation}
    {\rm{log}}(I_{\rm{R}}) = c + b \, {\rm{log}}(I_{\rm{X}}),
\end{equation}
where the slope $b$ quantifies the scaling between the thermal and non-thermal components of the ICM. This slope is related to the underlying particle acceleration mechanism responsible for the diffuse synchrotron emission \citep{Govoni2001,Brunetti2004,ZuHone2013,ZuHone2015}.

In general, a super-linear slope is expected in the secondary/hadronic scenario, due to the central CRp injection profile and relative scaling between the CR and thermal gas \citep[see e.g.][]{Ignesti2020}. In the primary/turbulent (re-)acceleration scenario, depending on the nature and distribution of CRe throughout the cluster volume, either a sub-linear or super-linear slope can be generated. 

We used \textsc{Linmix} \citep{Kelly2007} to determine the best-fitting values of $b$ and $c$. \textsc{Linmix} takes a hierarchical Bayesian approach to linear regression, accounting for uncertainties on both the independent and dependent variables --- in this case, $I_{\rm X}$ and $I_{\rm R}$ respectively --- as well as upper limits. This latter aspect is important, as the diffuse radio emission recovered by MeerKAT and LOFAR does not fill the entirety of the X-ray emitting volume of Abell~1413.

\textsc{Linmix} yields a best-fit slope of $b_{\rm{1283~MHz}} = 1.63^{+0.10}_{-0.10}$ and $b_{\rm{145~MHz}} = 1.20^{+0.13}_{-0.11}$. Note that we take the median value from our linear regression as the best-fit value, and use the 16th and 84th percentiles to define the uncertainty. Thus, there is clear evidence of a change in slope with frequency, at $\gtrsim 3.4 \sigma$ significance. There is no evidence of departure from a single correlation in the $I_{\rm R} / I_{\rm X}$ plane which might signify differences in the physical conditions and/or emission mechanism \citep[as in][]{Biava2021_RXCJ1720,Riseley2022_MS1455}. We summarise our fit results in Table~\ref{tab:correlation_results}.

As a further investigation, to account for the relative sensitivity of each of our datasets, we re-ran our \textsc{Linmix} fitting routine in the $I_{\rm{R}}/I_{\rm{X}}$ plane using a common set of regions where the radio surface brightness was measured to be in excess of $2\sigma$ at both 1283~MHz and 145~MHz. Regions where the radio surface brightness $I_{\rm{R}}$ was less than $2\sigma$ at either frequency were treated as upper limits in the same manner, adopting a value of $2\sigma$ in the respective map. We found a super-linear correlation with steeper slope of $b_{\rm{1283~MHz}} = 2.23^{+0.22}_{-0.19}$ and $b_{\rm{145~MHz}} = 1.36^{+0.15}_{-0.11}$, confirming the change in correlation slope.

\begin{figure}
\begin{center}
\includegraphics[width=0.99\linewidth]{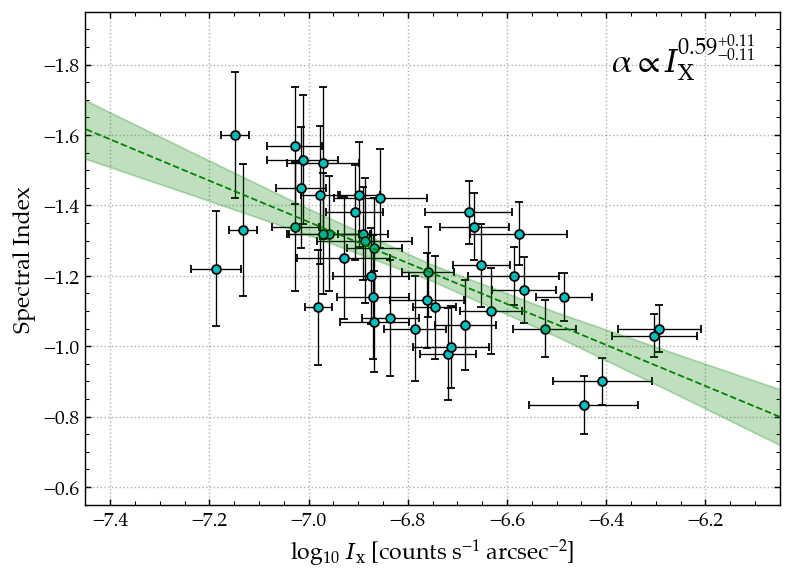}
\cprotect\caption{Radio spectral index/X-ray surface brightness correlation $(\alpha^{1283~{\rm{MHz}}}_{145~{\rm{MHz}}} / I_{\rm X})$ for the Abell~1413 mini-halo at 15~arcsec resolution. Note the inverted $y$-axis, to facilitate comparison with previous similar studies. Dashed green line shows the best-fit to Equation~\ref{eq:ptp_alfa}, derived using \textsc{Linmix}, and the shaded green region shows the $1\sigma$ uncertainty. The slope is $b = 0.59^{+0.11}_{-0.11}$.}
\label{fig:ptp_alfa}
\end{center}
\end{figure}

\begin{table}
\renewcommand{\arraystretch}{1.4}
\centering
\caption{Summary of results for our \textsc{Linmix} fitting routines, fitting the point-to-point correlation between X-ray surface brightness and either the radio surface brightness at the listed frequency or spectral index, as indicated in the first column. $b$ is the best-fit correlation slope for the plane, and $r_{\rm{S}}$ and $r_{\rm{P}}$ are respectively the Spearman and Pearson correlation coefficient for each plane. \label{tab:correlation_results}}
\begin{tabular}{lccc}
\hline
Image & Slope & Spearman coeff. & Pearson coeff. \\
      & $b$   & $r_{\rm{S}}$ & $r_{\rm{P}}$ \\
\hline\hline
1283~MHz  & $1.63^{+0.10}_{-0.10}$ &  0.90 & 0.94 \\
145~MHz   & $1.20^{+0.13}_{-0.11}$ &  0.84 & 0.86 \\
\hline
$\alpha$ & $0.59^{+0.11}_{-0.11}$ &  0.67 & 0.62 \\
\hline
\end{tabular}
\end{table}

\subsubsection{Point-to-point correlation: spectral index}
We also studied the spatial distribution of the radio spectral index, and the connection with the thermal properties of Abell~1413, via point-to-point analysis of the correlation between spectral index $\alpha$ and X-ray surface brightness, i.e. the $\alpha / I_{\rm X}$ plane.

Figure~\ref{fig:ptp_alfa} presents the $\alpha / I_{\rm X}$ correlation plane for Abell~1413, profiled using the same regions shown in Figure~\ref{fig:A1413_boxes}. Limits are more difficult to account for when considering the $\alpha / I_{\rm X}$ plane, as both upper and lower limits to the spectral index can be present depending on the relative sensitivity of the radio data and the underlying physical processes. See also discussion by \cite{Botteon2020b_Abell2255}.

As is visible in Figure~\ref{fig:A1413_alfa}, there are a number of regions toward the mini-halo outskirts where we cannot measure the spectral index, due to the limited sensitivity of the LOFAR data. While we could place a \emph{lower} limit on the spectral index for these regions, these are non-trivial to account for even with a Bayesian algorithm such as \textsc{Linmix}. Thus, we performed our point-to-point analysis using only regions where spectral index \emph{measurements} were possible: those regions above the $3\sigma$ level in both radio maps.

Figure~\ref{fig:ptp_alfa} appears to show a good correlation between $\alpha$ and $I_{\rm X}$; indeed, we find that the measurements are strongly correlated with correlation coefficients $r_{\rm S} = 0.67$ and $r_{\rm P} = 0.62$. Again, we used \textsc{Linmix} to perform the linear regression, fitting a power-law in log-linear space as follows:
\begin{equation}\label{eq:ptp_alfa}
    \alpha = c + b_{\alpha} \, {\rm{log}}(I_{\rm{X}})
\end{equation}

Our linear regression yields a best-fit slope of $b_{\alpha} = 0.59^{+0.11}_{-0.11}$; this slope is plotted in Figure~\ref{fig:ptp_alfa}, with the $1\sigma$ uncertainty traced by the shaded region. We observe no sign of departure from a single trend in the $\alpha / I_{\rm X}$ plane, in line with the observed spectral steepening in the spatially-resolved spectral index map presented in Figure~\ref{fig:A1413_alfa}.

\subsubsection{Point-to-point correlation: discussion}
Few mini-haloes in the literature have the necessary highly-sensitive multi-frequency radio data with which to examine both the $I_{\rm R} / I_{\rm X}$ and $\alpha / I_{\rm X}$ planes. Two examples are the mini-haloes hosted by RX~J1720.1$+$2638 and MS~1455.0$+$2232 (respectively \citealt{Biava2021_RXCJ1720}, \citealt{Riseley2022_MS1455}). Among a sample of seven mini-haloes, \cite{Ignesti2020} reports dual-frequency investigation of the $I_{\rm R} / I_{\rm X}$ for two clusters, Abell~3444 and 2A~0335$+$096, using historic narrow-band data; however, those authors do not study the $\alpha / I_{\rm X}$ correlation.

\cite{Biava2021_RXCJ1720} analyse the point-to-point correlations for the multiple diffuse components hosted by RX~J1720.1$+$2638: the known mini-halo and eastern extension, as well as the larger-scale diffuse emission seen only at LOFAR frequencies. These three components show clearly distinct trends, with the mini-halo showing a super-linear slope in the $I_{\rm R} / I_{\rm X}$ plane, and the others showing a sub-linear slope. In the $\alpha / I_{\rm X}$ plane, the mini-halo shows no correlation, whereas the other components each follow a single distinct correlation. For MS~1455.0$+$2232, the $I_{\rm R} / I_{\rm X}$ plane is characterised by a single correlation, whereas regions inside and outside the sloshing spiral showed different behaviour in the $\alpha / I_{\rm X}$ plane \citep[see][]{Riseley2022_MS1455}.

Aspects of our findings for Abell~1413 are similar. The correlation in the $\alpha / I_{\rm X}$ plane for Abell~1413 is similar to the behaviour outside the sloshing spiral in MS~1455.0$+$2232, where the correlation slope was $b_{\alpha} = 0.21 \pm 0.11$. Similarly, the single correlation in the $I_{\rm R} / I_{\rm X}$ plane is consistent with our findings for MS~1455.0$+$2232.

While the observed super-linear slope in the $I_{\rm R} / I_{\rm X}$ plane is typical of mini-haloes, it is not yet conclusive whether the frequency dependence of this slope is the norm. In this case of Abell~1413, the corrlation slope clearly changes with frequency. The mini-halo in 2A~0335$+$096 shows tentative evidence of a slope change with frequency, although the presence of sub-structures within the mini-halo means that the uncertainties are large \citep[see][]{Ignesti2020,Ignesti2021}. Abell~3444, RX~J1720.1$+$2638, and MS~1455.0$+$2232 show no change in the $I_{\rm R} / I_{\rm X}$ correlation slope with frequency \citep[respectively][]{Ignesti2020,Biava2021_RXCJ1720,Riseley2022_MS1455}.

Interpretation of the $I_{\rm R} / I_{\rm X}$ and $\alpha / I_{\rm X}$ correlations requires careful consideration. However, broadly-speaking the steepening of the $I_{\rm R} / I_{\rm X}$ correlation toward higher frequencies as well as the slope of the $\alpha / I_{\rm X}$ correlation both imply a steepening spectral index with increasing radius. Such a steepening would arise naturally in an acceleration scenario where turbulence plays a significant role in powering the diffuse emission, as has been seen in some radio haloes \citep[e.g.][]{Rajpurohit2021_MACSJ0717_Halo}. However, this is the first clear confirmation of a radially-steepening mini-halo spectrum.

\section{On the nature of the mechanism powering the mini-halo in Abell~1413}
Our census has the over-arching aim of investigating and understanding the underlying mechanism responsible for generating mini-haloes. We aim to answer the question of whether the secondary electron model (hadronic collisions between relativistic protons and thermal protons generating CRe) or the primary electron model (acceleration of electrons to relativistic energies by cluster-scale turbulence) is responsible, or whether some form of hybrid scenario predicts properties that are more consistent with our observations.

Similar to our previous study of MS~1455.0$+$2232, the evidence accumulated from our study of Abell~1413 is mixed. The super-linear $I_{\rm R} / I_{\rm X}$ correlation arises naturally within the secondary electron model, although the primary model can also replicate a super-linear slope depending on the nature of the turbulence. 

The numerous active radio galaxies associated with Abell~1413, which are distributed across much of the cluster volume, would naturally provide a source of relativistic protons responsible for generation of CRe via the secondary electron model.

On the other hand the asymmetry of the mini-halo, which is elongated in the north/south direction, and evidence of large-scale disruption in the ICM --- the disturbed BCG, substructure in the galaxy distribution, elongated X-ray surface brightness distribution, and `warm core' --- all suggest that turbulence is likely present on large scales in the ICM. Similarly, the non-uniformity of the spectral index and the tentative evidence of multiple components in the mini-halo would arise more naturally under the primary model, as turbulence is inherently an intermittent process in both spatial and temporal terms.

Overall, we cannot form strong conclusions on the mechanism responsible for the generation of the mini-halo in Abell~1413. Both mechanisms are able to replicate some of the observed properties, but neither mechanism reproduces all of them comprehensively. While on balance the evidence somewhat favours an interpretation of the turbulent acceleration framework, it is likely that both mechanisms are active to some extent.

\subsection{A simple mathematical framework for hybrid models}
Hybrid models have been previously proposed by several authors in the past, predominantly in the context of radio haloes \citep[e.g.][]{Brunetti2005,Brunetti2011,Cassano2012,Zandanel2014,Pinzke2017}. Specifically, the models presented by \cite{Brunetti2005} and \cite{Brunetti2011} discuss the models most relevant in the context of our census: CRp and their secondary electrons undergoing re-acceleration by turbulence. In light of our observational results it is prudent to further the investigation of hybrid models and present a simple mathematical framework for the expected point-to-point correlation slope in this scenario.

To determine the correlation slope predicted by a hybrid scenario, we need to determine the predicted relation between the X-ray emissivity $\epsilon_{\rm X}$ and the radio emissivity $\epsilon_{\rm R}$. These two quantities take the following form:
\begin{subequations}
    \begin{equation}\label{eq:hybrid_xray}
        \epsilon_{\rm X} \propto n^2_{\rm ICM}
    \end{equation}
    \begin{equation}\label{eq:hybrid_radio}
        \epsilon_{\rm R} \propto F_{\rm t} \, \eta_{\rm e} \,  \frac{B^2}{B^2 + B^2_{\rm IC}}
    \end{equation}
\end{subequations}
where $n_{\rm ICM}$ is the number density of thermal particles in the ICM, $B$ is the magnetic field strength and $B_{\rm IC}$ is the inverse-Compton magnetic field strength \citep[e.g.][]{Brunetti_Vazza_2020_PRL}. $F_{\rm t}$ is the turbulent flux defined as:
\begin{equation}\nonumber\label{eq:hybrid_def1}
    F_{\rm t} \propto n_{\rm ICM} \, \frac{\delta V^3_{\rm t}}{L_{\rm t}}
\end{equation}
where $L_{\rm t}$ is the turbulence length scale and $\delta V^3_{\rm t}$ is the dispersion of the turbulent velocity field on that scale. During cluster mergers and interactions, it is generally expected that the turbulence occurs on scales of the cluster core, i.e. $L_{\rm t} \simeq 0.1$~to~$0.4$~Mpc \citep[e.g.][]{Vazza2009}. The acceleration efficiency $\eta_{\rm e}$ is defined as:
\begin{equation}\nonumber\label{eq:hybrid_def1}
    \eta_{\rm e} \propto F^{-1}_{\rm t} \int {\rm d}^3 p \, \frac{E}{p^2} \, \frac{\partial}{\partial p} \left( p^2 D_{pp} \frac{\partial f}{\partial p} \right) \simeq \frac{U_{\rm CRe}}{F_{\rm t}} \frac{D_{pp}}{p^2}
\end{equation}
where $U_{\rm CRe}$ is the energy density of CRe, $p$ is the particle momentum and $D_{pp}$ is the momentum diffusion coefficient \citep[e.g.][and references therein]{Brunetti_Lazarian_2007,BrunettiJones2014}. Substituting these definitions into Equation~\ref{eq:hybrid_radio} we derive:
\begin{equation}\label{eq:hybrid_radio2}
    \epsilon_{\rm R} \propto U_{\rm CRe} \, \frac{D_{pp}}{p^2} \,  \frac{B^2}{B^2 + B^2_{\rm IC}}
\end{equation}

Taking a magnetic field strength that follows the relation $B^2 \propto n_{\rm ICM}$ and using Equation~\ref{eq:hybrid_xray} we find that:
\begin{equation}\label{eq:hybrid_radio3}
    \epsilon_{\rm R} \propto U_{\rm CRe} \, \frac{D_{pp}}{p^2} \, \frac{ 1 }{ 1 + \left( \frac{B_{\rm IC}}{B0}^2 \right) \left( \epsilon_{\rm X} \right)^{-1/2} }
\end{equation}
where $B_0$ is the central magnetic field strength. In the case where we have a hybrid model, we have a CRe energy density $U_{\rm CRe}$ that is a function of both the ICM number density $n_{\rm ICM}$ and the number density of cosmic ray protons $n_{\rm CRp}$:
\begin{equation}\label{eq:hybrid_ucre}
    U_{\rm CRe} \propto n_{\rm ICM} \, n_{\rm CRp}
\end{equation}
and hence Equation~\ref{eq:hybrid_radio3} becomes:
\begin{equation}\label{eq:hybrid_radio4}
    \epsilon_{\rm R} \propto \left( n_{\rm CRp} \frac{D_{pp}}{p^2} \right) \, \frac{ \epsilon_{\rm X}^{1/2} }{ 1 + \left( \frac{B_{\rm IC}}{B0}^2 \right) \left( \epsilon_{\rm X} \right)^{-1/2} }
\end{equation}
If we assume the CRp population in the ICM is dominated by injection from the AGN of the central radio BCG, and assume for simplicity a spatial diffusion coefficient $\kappa_0$ which is both energy-independent and constant with respect to distance from the injection source, the CRp number density in the ICM $n_{\rm CRp}$ is given by:
\begin{equation}\nonumber\label{eq:hybrid_ncrp}
    n_{\rm CRp} \propto \frac{ Q_{\rm CRp} }{ \kappa_0 \, r }
\end{equation}
where $r$ is the radius and $Q_{\rm CRp}$ is the injection rate from the radio BCG. Substituting this definition into Equation~\ref{eq:hybrid_radio4} and normalising by the central values of $\epsilon_{\rm R}$ and $\epsilon_{\rm X}$, respectively $\epsilon_{\rm R,0}$ and $\epsilon_{\rm X,0}$ we find that:
\begin{equation}\label{eq:hybrid_radio5}
    \frac{\epsilon_{\rm R}}{\epsilon_{\rm R,0}} \propto \frac{1}{r} \, \left( \frac{ Q_{\rm CRp} }{ \kappa_0 } \, \frac{D_{pp}}{p^2} \right) \, \left( \frac{ \left( \epsilon_{\rm X} / \epsilon_{\rm X,0} \right)^{1/2} }{ 1 + \left( \frac{B_{\rm IC}}{B0}^2 \right) \left( \epsilon_{\rm X} / \epsilon_{\rm X,0} \right)^{-1/2} } \right)
\end{equation}
where the quantity inside the first set of brackets is constant, and in the case where $B_0^2 \gg B_{\rm IC}^2$, we find that:
\begin{equation}\label{eq:hybrid_radiofinal}
    \frac{\epsilon_{\rm R}}{\epsilon_{\rm R,0}} \propto \frac{1}{r} \, \left( \frac{ \epsilon_{\rm X} }{  \epsilon_{\rm X,0} } \right)
\end{equation}
and hence Equation~\ref{eq:hybrid_radiofinal} indicates that the hybrid scenario predicts a \textit{super}-linear correlation between the radio and X-ray emissivities, and therefore in the $I_{\rm R} / I_{\rm X}$ plane. This simple mathematical framework suggests that the super-linear point-to-point correlation slope reported here for Abell~1413 and previously for MS~1455.0$+$2232 \citep{Riseley2022_MS1455} are both compatible with a hybrid scenario.

\section{Conclusions}\label{sec:conclusions}
This paper is the second in a series of papers presenting the results from a `MeerKAT-meets-LOFAR' mini-halo census, covering 13 clusters hosting known mini-haloes that are visible to both MeerKAT and LOFAR for long-track observations (typically 5.5~hours on-source per observing run).

We have presented new MeerKAT L-band (1283~MHz) and LOFAR HBA (145~MHz) observations of the galaxy cluster Abell~1413, which hosts a known mini-halo. We have combined our radio data with archival \textit{Chandra} observations, enabling us to perform a detailed comparison of the thermal and non-thermal properties of this intriguing cluster.

Our new, deep radio observations allow us to achieve more sensitive results than previous studies of this cluster. At full resolution we detect many compact radio sources in the vicinity of Abell~1413 as well as several tailed radio galaxies that are either likely or confirmed cluster-members. After subtracting the contaminating sources and tapering our data to enhance our sensitivity to diffuse radio emission, we detect faint and highly-extended radio emission from the `mini'-halo up to $\sim 584$~kpc (and at least 449~kpc) at 1283~MHz.

Measuring from our source-subtracted maps at 15~arcsec resolution and integrating over the region where we are most confident in our source subtraction, we derive a total integrated flux density of $S_{\rm 1283~MHz} = 3.23 \pm 0.17$~mJy and $S_{\rm 145~MHz} = 29.5 \pm 3.4$~mJy. Using these values, we measure an integrated spectral index of $\alpha_{\rm 145~MHz}^{\rm 1283~MHz} = -1.01 \pm 0.06$, and derive a $k$-corrected 1.4~GHz radio power of $P_{\rm 1.4~GHz} = (1.50 \pm 0.08) \times 10^{23}$~W~Hz$^{-1}$.

Using our exquisite radio data, we have examined the spatially-resolved spectral index profile of the mini-halo. We find an overall global median of $\langle \alpha \rangle = -1.18 \pm 0.11$. However, we also find tentative evidence of two different trends: the inner region of the mini-halo appears to show a slightly flatter spectrum with a median value $\langle \alpha_{\rm in} \rangle = -0.97 \pm 0.07$, whereas the outer regions of the mini-halo show a steeper value of $\langle \alpha_{\rm out} \rangle = -1.12 \pm 0.13$. This implies spectral steepening. We emphasise however that this is inconsistent only at the $\sim1\sigma$ level, and deeper low-frequency observations would be required to study this further and clarify the underlying emission mechanism.

We have studied the point-to-point correlations between X-ray surface brightness and (i) radio surface brightness at both 1283~MHz and 145~MHz (the $I_{\rm R} / I_{\rm X}$ correlation) and (ii) radio spectral index (the $\alpha / I_{\rm X}$ correlation). Our investigation shows that the radio/X-ray surface brightness is strongly correlated, with coefficients in the range $r = 0.84$ to $r = 0.94$ depending on frequency and type of coefficient. We find no evidence of departure from a single correlation.

At both frequencies considered here, the slope of the $I_{\rm R} / I_{\rm X}$ correlation is positive, with a value of $b = 1.63^{+0.10}_{-0.10}$ at 1283~MHz and $b = 1.20^{+0.13}_{-0.11}$ at 145~MHz. This change in slope with observing frequency may indicate a difference in the non-thermal/thermal connection, although the exact cause of this is as-yet unknown; it may reflect changes in the acceleration mechanism, ambient medium, and/or projection effects. In exploring the $\alpha / I_{\rm X}$ correlation, we find a clear correlation with a moderate strength (coefficients $0.62$ and $0.67$). The slope of this correlation is positive, with a value of $b_{\alpha} = 0.59^{+0.11}_{-0.11}$. 

While the super-linear slope is a typical signature of the hadronic scenario, our observations also support the interpretation that there is large-scale turbulence at work in Abell~1413. Hence, we investigated a simple mathematical framework which demonstrates that hybrid models --- whereby secondary electrons are re-accelerated by turbulence --- naturally reproduce a super-linear correlation slope in the $I_{\rm R} / I_{\rm X}$ plane.

\section*{Acknowledgements}
CJR, NB, and A.~Bonafede acknowledge financial support from the ERC Starting Grant `DRANOEL', number 714245. EB acknowledges support from DFG FOR5195. KR acknowledges funding from \textit{Chandra} grant GO0-21112X and ERC starting grant `MAGCOW' number 714196. FL acknowledges financial support from the Italian Minister for Research and Education (MIUR), project FARE, project code R16PR59747, project name FORNAX-B. FL acknowledges financial support from the Italian Ministry of University and Research $-$ Project Proposal CIR01$\_$00010. RT and RJvW acknowledge support from the ERC Starting Grant `ClusterWeb', number 804208. EO and RJvW acknowledge support from the VIDI research programme with project number 639.042.729, which is financed by the Netherlands Organisation for Scientific Research (NWO). We thank our anonymous referee for their feedback on our manuscript,  their constructive comments and their raising of interesting discussion points, which improved the quality of our publication.

The MeerKAT telescope is operated by the South African Radio Astronomy Observatory, which is a facility of the National Research Foundation, an agency of the Department of Science and Innovation. We wish to acknowledge the assistance of the MeerKAT science operations team in both preparing for and executing the observations that have made our census possible.

LOFAR is the Low Frequency Array designed and constructed by ASTRON. It has observing, data processing, and data storage facilities in several countries, which are owned by various parties (each with their own funding sources), and which are collectively operated by the ILT foundation under a joint scientific policy. The ILT resources have benefited from the following recent major funding sources: CNRS-INSU, Observatoire de Paris and Universit\'e d'Orl\'eans, France; BMBF, MIWF-NRW, MPG, Germany; Science Foundation Ireland (SFI), Department of Business, Enterprise and Innovation (DBEI), Ireland; NWO, The Netherlands; The Science and Technology Facilities Council, UK; Ministry of Science and Higher Education, Poland; The Istituto Nazionale di Astrofisica (INAF), Italy.

This research made use of the Dutch national e-infrastructure with support of the SURF Cooperative (e-infra 180169) and the LOFAR e-infra group. This work is co-funded by the EGI-ACE project (Horizon 2020) under Grant number 101017567. The J\"{u}lich LOFAR Long Term Archive and the German LOFAR network are both coordinated and operated by the J\"{u}lich Supercomputing Centre (JSC), and computing resources on the supercomputer JUWELS at JSC were provided by the Gauss Centre for Supercomputing e.V. (grant CHTB00) through the John von Neumann Institute for Computing (NIC). 

This research made use of the University of Hertfordshire high-performance computing facility and the LOFAR-UK computing facility located at the University of Hertfordshire and supported by STFC [ST/P000096/1], and of the Italian LOFAR IT computing infrastructure supported and operated by INAF, and by the Physics Department of Turin university (under an agreement with Consorzio Interuniversitario per la Fisica Spaziale) at the C3S Supercomputing Centre, Italy.

Funding for the Sloan Digital Sky Survey IV has been provided by the Alfred P. Sloan Foundation, the U.S. Department of Energy Office of Science, and the Participating Institutions. SDSS-IV acknowledges support and resources from the Center for High Performance Computing at the University of Utah. The SDSS website is \url{www.sdss.org}.

SDSS-IV is managed by the Astrophysical Research Consortium for the Participating Institutions of the SDSS Collaboration including the Brazilian Participation Group, the Carnegie Institution for Science, Carnegie Mellon University, Center for Astrophysics | Harvard \& Smithsonian, the Chilean Participation Group, the French Participation Group, Instituto de Astrof\'isica de Canarias, The Johns Hopkins University, Kavli Institute for the 
Physics and Mathematics of the Universe (IPMU) / University of Tokyo, the Korean Participation Group, Lawrence Berkeley National Laboratory, Leibniz Institut f\"ur Astrophysik Potsdam (AIP), Max-Planck-Institut f\"ur Astronomie (MPIA Heidelberg), Max-Planck-Institut f\"ur Astrophysik (MPA Garching), Max-Planck-Institut f\"ur Extraterrestrische Physik (MPE), National Astronomical Observatories of China, New Mexico State University, New York University, University of Notre Dame, Observat\'ario Nacional / MCTI, The Ohio State University, Pennsylvania State University, Shanghai Astronomical Observatory, United Kingdom Participation Group, Universidad Nacional Aut\'onoma de M\'exico, University of Arizona, University of Colorado Boulder, University of Oxford, University of Portsmouth, University of Utah, University of Virginia, University of Washington, University of Wisconsin, Vanderbilt University, and Yale University.

Finally, we acknowledge the developers of the following python packages (not mentioned explicitly in the text), which were used extensively during this project: \textsc{aplpy} \citep{Robitaille2012}, \textsc{astropy} \citep{Astropy2013}, \textsc{cmasher} \citep{vanderVelden2020}, \textsc{colorcet} \citep{Kovesi2015}, \textsc{matplotlib} \citep{Hunter2007}, \textsc{numpy} \citep{Numpy2011} and \textsc{scipy} \citep{Jones2001}.

\section*{Data Availability}
The images underlying this article will be shared on reasonable request to the corresponding author. Raw MeerKAT visibilities for both projects SCI-20210212-CR-01 and `Mining Minihalos with MeerKAT' are in the public domain. All MeerKAT data are accessed via the SARAO archive (\url{https://apps.sarao.ac.za/katpaws/archive-search}). Raw LOFAR visibilities can be accessed via the LOFAR Long-Term Archive (LTA; \url{https://lta.lofar.eu}). \textit{Chandra} data are available via the Chandra Data Archive (\url{https://cxc.harvard.edu/cda/}).

\bibliographystyle{mnras}
\bibliography{A1413_MeerKAT-meets-LOFAR}

\appendix

\begin{table*}
\renewcommand\thetable{A1}
\centering
\caption{Flux density measurements $S$ at 1283~MHz and 145~MHz, as well as the corresponding radio spectral index $\alpha$ for sources identified with yellow `+' signs in Figure~2 of the main manuscript. Point sources were modelled using a simple single Gaussian fit; extended source flux densities were derived by integrating above the $3\sigma_{\rm local}$ level. We note that for clarity the marker for source 14, the embedded head-tail radio galaxy, is not shown in Figure~2. This source, along with the BCG can be seen in Figure~4; the flux density measurements for the BCG are presented in Table~2 and Figure~5. Cross-identifications with Sloan Digital Sky Survey (SDSS) optical counterparts are listed where available, along with redshift measurements ($z$). We quote spectroscopic redshifts where present, photometric redshifts otherwise (identified with a `$p$'). Galaxies where the association and/or the SDSS photometry is uncertain are marked by a `$u$'. \label{tab:measurement}}
\begin{tabular}{lrrcccrlc}
\hline
ID & Right Ascension & Declination & $S_{\rm 1283~MHz}$ & $S_{\rm 145~MHz}$ & $\alpha$ & Cross-ID & $z$ & Notes \\
   & (J2000) & (J2000) & [mJy] & [mJy] & & & & \\
\hline\hline 
01  & 11:55:08.94 & +23:26:22.6  & $26.63 \pm 1.33$ & $178 \pm 18$ & $-0.87 \pm 0.05$ & SDSS~J115508.97$+$232623.4 & 0.144 & $-$ \\
02  & 11:55:12.73 & +23:22:53.9  & $0.049 \pm 0.006$ & $-$ & $-$ & SDSS~J115512.73$+$232254.3 & $-$ & $-$ \\
03  & 11:55:13.37 & +23:24:29.2  & $0.090 \pm 0.007$ & $-$ & $-$ & $-$ & $-$ & $-$ \\
04  & 11:55:13.49 & +23:22:47.5  & $0.053 \pm 0.005$ & $-$ & $-$ & $-$ & $-$ & $-$ \\
05  & 11:55:13.87 & +23:23:27.1  & $0.039 \pm 0.004$ & $-$ & $-$ & $-$ & $-$ & $-$ \\
06  & 11:55:14.00 & +23:26:31.1  & $0.110 \pm 0.022$ & $1.11 \pm 0.13$ & $-1.06 \pm 0.11$ & SDSS~J115513.96$+$232630.7 & 0.171 & $p$\\
07  & 11:55:14.82 & +23:26:37.9  & $0.213 \pm 0.017$ & $0.98 \pm 0.13$ & $-0.70 \pm 0.07$ & SDSS~J115515.46$+$232635.1 & 0.666 & $p,u$ \\
08  & 11:55:15.11 & +23:24:01.4  & $0.46 \pm 0.02$ & $1.10 \pm 0.12$ & $-0.40 \pm 0.05$ & $-$ & $-$ & $-$ \\
09  & 11:55:15.47 & +23:25:39.9  & $0.19 \pm 0.02$ & $0.68 \pm 0.39$ & $-0.58 \pm 0.27$ & SDSS~J115515.47$+$232539.6 & 0.176 & $p$ \\
10  & 11:55:15.92 & +23:26:25.2  & $0.294 \pm 0.021$ & $1.37 \pm 0.26$ & $-0.71 \pm 0.09$ & SDSS~J115516.27$+$232627.2 & $-$ & $u$ \\
11  & 11:55:16.54 & +23:21:43.6  & $0.063 \pm 0.005$ & $-$ & $-$ & SDSS~J115516.62$+$232142.6 & $-$ & $-$ \\
12  & 11:55:17.06 & +23:23:55.1  & $0.10 \pm 0.02$ & $-$ & $-$ & SDSS~J115517.13$+$232352.6 & 0.149 & $p$ \\
13  & 11:55:17.12 & +23:22:14.7  & $1.89 \pm 0.10$ & $16.42 \pm 1.75$ & $-0.99 \pm 0.05$ & SDSS~J115518.60$+$232424.0 & 0.139 & $-$ \\
14  & 11:55:18.64 & +23:24:21.2  & $2.39 \pm 0.12$ & $12.16 \pm 1.33$ & $-0.75 \pm 0.06$ & SDSS~J115518.60$+$232424.0 & 0.139 & $-$ \\
15  & 11:55:18.56 & +23:22:03.7  & $0.058 \pm 0.006$ & $-$ & $-$ & SDSS~J115521.05$+$232319.2 & 0.351 & $p,u$ \\
16  & 11:55:19.26 & +23:26:51.3  & $0.330 \pm 0.017$ & $3.56 \pm 0.45$ & $-1.09 \pm 0.06$ & SDSS~J115519.28$+$232651.6 & $0.464$ & $p$ \\
17  & 11:55:20.84 & +23:23:18.9  & $0.051 \pm 0.006$ & $-$ & $-$ & $-$ & $-$ & $-$ \\
18  & 11:55:21.12 & +23:27:24.3  & $0.108 \pm 0.007$ & $0.86 \pm 0.10$ & $-0.95 \pm 0.06$ & SDSS~J115521.09$+$232723.6 & 0.289 & $p$ \\
19  & 11:55:21.42 & +23:23:36.1  & $0.12 \pm 0.02$ & $1.04 \pm 0.13$ & $-0.99 \pm 0.10$ & SDSS~J115521.97$+$232335.5 & 0.373 & $u$ \\
20  & 11:55:22.16 & +23:26:18.8  & $0.086 \pm 0.006$ & $0.51 \pm 0.06$ & $-0.82 \pm 0.06$ & $-$ & $-$ & $-$ \\
21  & 11:55:23.04 & +23:23:04.4  & $0.25 \pm 0.01$ & $-$ & $-$ & $-$ & $-$ & $-$ \\
22  & 11:55:23.17 & +23:24:25.0  & $0.049 \pm 0.005$ & $0.85 \pm 0.13$ & $-1.31 \pm 0.08$ & SDSS~J115523.03$+$232425.4 & 0.140 & $u$ \\
23  & 11:55:23.83 & +23:26:59.5  & $0.071 \pm 0.013$ & $-$ & $-$ & SDSS~J115524.30$+$232654.4 & 0.706 & $p,u$ \\
24  & 11:55:24.15 & +23:24:43.0  & $0.21 \pm 0.02$ & $0.91 \pm 0.39$ & $-0.67 \pm 0.20$ & SDSS~J115524.16$+$232443.0 & 1.775 & $p,u$ \\
25  & 11:55:25.13 & +23:26:40.4  & $0.080 \pm 0.005$ & $-$ & $-$ & SDSS~J115525.19$+$232640.9 & 0.770 & $p$ \\
\hline
\end{tabular}
\end{table*}

\bsp	
\label{lastpage}
\end{document}